\documentclass{article}
\usepackage[a4paper, showframe=False]{geometry}
\usepackage{xcolor}
\definecolor{kitfarbe}{rgb}{0.004,0.588,0.51}
\definecolor{urlcol}{HTML}{294E85}
\definecolor{anccol}{HTML}{B52D26}  

\usepackage[utf8]{inputenc} 
\usepackage[T1]{fontenc}    
\usepackage[colorlinks=true,
            urlcolor=urlcol,
            linkcolor=anccol,
            citecolor=kitfarbe]{hyperref}   

\usepackage{svg}
\usepackage{lmodern,textcomp}
\usepackage{makecell}
\usepackage{booktabs}
\usepackage{natbib}
\usepackage{amsfonts}       
\usepackage{nicefrac}       
\usepackage{microtype}      
\usepackage{tabularx}
\usepackage{adjustbox}
\usepackage{soul}
\usepackage{placeins}

\usepackage{siunitx}
\usepackage{amsmath}
\usepackage{mathtools}
\usepackage{cancel}
\usepackage{amsthm}
\usepackage{amssymb}
\usepackage{bbm}
\usepackage{bm}
\usepackage{subcaption}
\usepackage{tikz}

\usepackage{caption}

\newcommand{\Cov}{\mathrm{Cov}}
\newcommand{\E}{\mathbb{E}}
\def\gB{{\mathcal{B}}}
\def\gL{{\mathcal{L}}}
\def\gR{{\mathcal{R}}}
\def\gT{{\mathcal{T}}}
\def\rvu{{\mathbf{u}}}
\def\rvom{{\mathbf{\omega}}}

\def\sR{{\mathbb{R}}}
\newcommand{\train}{\mathcal{D}}
\newcommand{\Var}{\mathrm{Var}}
\def\vx{{\bm{x}}}

\def\ind{\mathbbm{1}}

\usepackage[utf8]{inputenc}
\title{Simplifying Random Forests' \\ Probabilistic Forecasts\thanks{We thank conference participants at the International Symposium on Forecasting (Dijon, 2024) as well as Andreas Eberl and two anonymous reviewers for helpful comments.}}

\author{%
Nils Koster \\
Institute of Statistics, Karlsruhe Institute of Technology, \\
Broad Institute of MIT and Harvard \\
\texttt{nils.koster@kit.edu} \\
\and
Fabian Kr\"uger \\
Institute of Statistics, Karlsruhe Institute of Technology \\
\texttt{fabian.krueger@kit.edu} \\
}

\date{\today}

\begin{document}

\maketitle

\begin{abstract}
Since their introduction by \citet{Breiman2001}, Random Forests (RFs) have proven to be useful for both classification and regression tasks. 
The RF prediction of a previously unseen observation can be represented as a weighted sum of all training sample observations.
This nearest-neighbor-type representation is useful, among other things, for constructing forecast distributions \citep{Meinshausen2006}. 
In this paper, we consider simplifying RF-based forecast distributions by sparsifying them. That is, we focus on a small subset of $k$ nearest neighbors while setting the remaining weights to zero.
This simplification, which we refer to as `Top$k$', greatly improves the interpretability of RF predictions. It can be applied to any forecasting task without re-training existing RF models. In empirical experiments, we document that the simplified predictions can be similar to or exceed the original ones in terms of forecasting performance. We explore the statistical sources of this finding via a stylized analytical model of RFs. 
The model suggests that simplification is particularly promising if the unknown true forecast distribution contains many small weights that are estimated imprecisely. 
\end{abstract}


\section{Introduction}
\label{sec:intro}
Many statisticians agree that forecast distributions are preferable to mere point forecasts. Correspondingly, an active literature is concerned with making statistical forecast distributions in meteorology \citep[e.g.][]{RaspLerch2018}, economics \citep[e.g.][]{KruegerEtAl2017}, energy \citep[e.g.][]{TaiebEtAl2021}, epidemiology \citep[e.g.][]{CramerEtAl2022}, and other fields. 
Nevertheless, point predictions still dominate in many practical settings in policy, business, and society. 
As argued by \citet{Raftery2016}, the cognitive load that forecast distributions impose upon their users may be an important bottleneck impeding their adoption. 
Motivated by this possibility, we consider simplifying the forecast distributions produced by Random Forests \citep[RFs;][]{Breiman2001,Meinshausen2006}, and study how simplification affects statistical forecasting performance. 
While our main focus is on probabilistic forecasting, the method we propose can also be used to simplify point forecasts for the mean. 

\begin{figure}
    \centering
    \includegraphics[width=1.\textwidth]{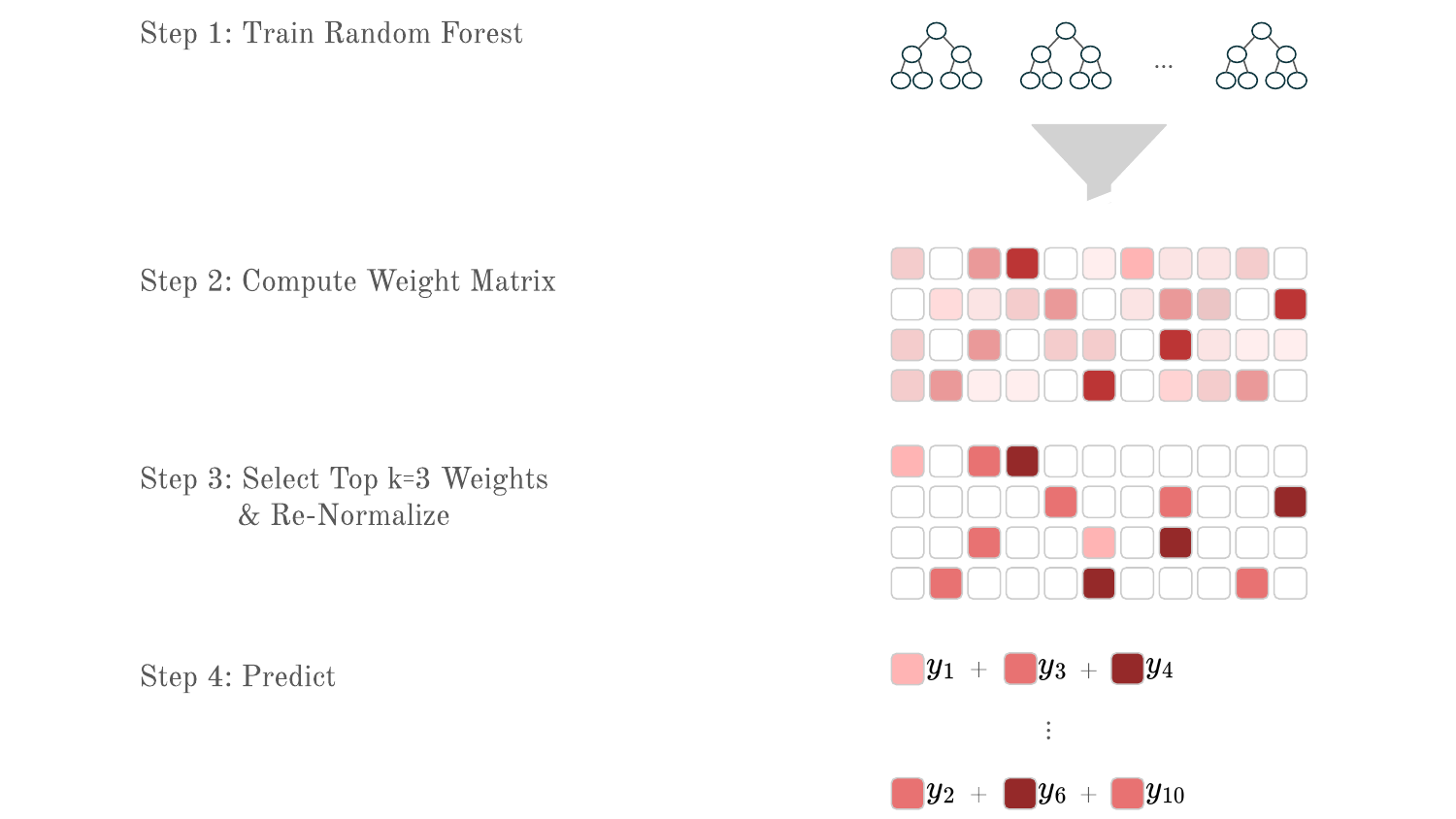}
    \caption{\textbf{Summary of proposed Top$k$ method.} This schematic description summarizes the method proposed in this paper. First, a standard RF is trained. Second, from the trained RF, we compute a weight vector for each new test case. Collecting the weight vectors for multiple test cases results in a matrix, with rows corresponding to test cases and columns corresponding to training cases. Next, we select the top $k$ (in this case $k=3$) weights for each test case, re-normalize such that the weights again sum up to one and set the remaining weights to zero. These weights can now be used for prediction, as illustrated here for the first and fourth test cases.}
    \label{fig:schematic_method}
\end{figure}

More specifically, we approximate an RF forecast distribution for a continuous scalar outcome by a discrete distribution with $k$ support points, where $k \in \{1, 2, \ldots, n\}$ is a user-determined parameter and $n$ denotes the number of training samples. 
The resulting forecast distribution can be cast as a collection of $k$ scenarios, each occurring with a specified probability. 
If $k$ is sufficiently small, this setup can effectively be communicated to non-statisticians. 
For example, \cite{AltigEtAl2022} use it to survey business executives about firm outcomes like future sales growth and employment. 
\citet[Chapter 35]{AbbasHoward2015} provide a textbook discussion from a decision analysis perspective.

Our approximation, which we refer to as `Top$k$' in the following, builds upon the fact that RFs can be cast as a nearest-neighbor-like method \citep{Lin2006, Meinshausen2006}: 
The forecast distribution for a test sample observation with feature vector $\vx_0$ is a discrete distribution with support points given by the training sample outcomes $(y_i)_{i=1}^n$ and corresponding probability weights $(w_i(\vx_0))_{i=1}^n$. \cite{Meinshausen2006} called these forecast distributions Quantile Regression Forests (QRFs). In the following, we do not distinguish between QRFs and RFs since the former can be constructed from the latter without any additional model fitting. As detailed in Section \ref{subsec:theory_background}, the weight $w_i(\vx_0)$ reflects the similarity between $\vx_0$ and $\vx_i$, the feature vector of the $i$-th training sample observation. Consider a given test case $\vx_0$ and the set $\mathcal{I}_{k}(\vx_0)$ containing the indices of the $k$ largest weights for this test case, where $k \in \{1, 2, \ldots, n\}$. 
We then set
\begin{equation*}
    \tilde{w}_i(\vx_0) = 
    \begin{cases} 
        \frac{w_i(\vx_0)}{\sum_{j\in\mathcal{I}_k(\vx_0)} w_j(\vx_0)} &~\text{if}~i \in \mathcal{I}_k(\vx_0)\\ 
        0& ~\text{else}
    \end{cases} 
\end{equation*}
That is, we retain only the $k << n$ largest weights, and re-scale them such that they sum to one. 
All other weights are set to zero. Figure \ref{fig:schematic_method} provides a schematic description. Setting $k = n$ recovers the weights of the initial RF's mean and the QRF's distribution forecast, which is described in more detail in Section \ref{sec:theory}.\footnote{Prediction as in Equations \ref{eq:rf_weights} and \ref{eq:rf_ecdf_pred} is unaltered, except for the substitution of $w_i(\vx_0)$ by $\tilde{w}_i(\vx_0)$.} 
Larger values of $k$ correspond to a larger number of scenarios. 
This increases the complexity of the forecast distribution, but may be beneficial in terms of statistical forecasting performance. 

\paragraph{Practical Illustration}
We provide a practical illustration of our method by considering labor market survey data provided by \cite{SoepPractice}.
The data set is an openly available subsample of the German Socioeconomic Panel \citep[SOEP;][]{GoebelEtAl2019}. 
It is recorded between 2015 and 2019, covering \num{21851} observations corresponding to \num{5847} distinct persons after preprocessing. Many participants were asked repeatedly over the survey years, so that the data has a panel structure.
Our goal here is to predict a person's annual salary in Euros based on socioeconomic regressors such as age, education, family status, and the industry in which the person is employed. 
We use data from 2015--2018 for training and 2019 for testing, and consider a Top$k$ model with $k=5$ along with a full RF. 
Compared to the full RF, the Top$k$ model is about 6\% worse in terms of distribution forecasting performance (as measured by the CRPS, introduced in Section \ref{sec:crps}), and about 3\% better in terms of point forecasting performance (as measured by the squared error). 
Section A of the online supplement provides details on the data set and RF forecasting performance.

To illustrate, consider a fully employed 47-year-old male, living alone and working in the financial sector.
We aim to predict his salary in 2019. 
A full RF predicts a mean income of \num{84705}, placing non-zero weight on 171 training sample observations. 
The simplified Top5 model utilizes only the five largest of these 171 weights which (before re-normalization) sum up to 0.47. 
To re-normalize, we simply multiply the five remaining weights by $1/0.47$. 
The five support points (the training sample observations that correspond to the remaining weights) are listed in Table \ref{tab:soep_ex}, together with their weights. 
With those, the Top5 model predicts a mean income of \num{84184}. 

\vspace{.5cm}
\begin{table}[!h]
	\centering
	\small
	\caption{\textbf{SOEP test prediction.} 
		The table shows the five support points $\vx_{s1}$ to $\vx_{s5}$ for a test point (first row, $\vx_0$) in the Top5 model. 
		`Empl.' abbreviates the variable `employed', `F' abbreviates `full employment'. 
		Sector ID 64 corresponds to `Provision of financial services'.
		The corresponding (rounded) weight of each support point is depicted in the penultimate column, while we report the respective income in the last column.
	}
	\label{tab:soep_ex}
	\begin{tabular}{lllllllll@{\hskip 0.2in}rr}
		\toprule
		& Survey  & Female & Age  & No.     & No.             & Years   & Empl.  & Sector & $w_i(\vx_0)$ [\%] & Income \\
		& Year    &        &      & Per.    & Child.          & Educ.   &        & ID     &              & \\
		\midrule
		$\vx_0$ & 2019 & False & 47 & 1 & 0 & 14.5 & F & 64 & - & \num{83279} \\
		\midrule
		$\vx_{s1}$ & 2017 & False & 45 & 1 & 0 & 14.5 & F & 64 & 30.6 & \num{94903} \\
		$\vx_{s2}$ & 2018 & False & 45 & 1 & 0 & 18   & F & 64 & 24.6 & \num{79206} \\
		$\vx_{s3}$ & 2017 & False & 44 & 1 & 0 & 15   & F & 64 & 17.3 & \num{100433} \\
		$\vx_{s4}$ & 2016 & False & 44 & 1 & 0 & 14.5 & F & 64 & 16.0 & \num{51608} \\
		$\vx_{s5}$ & 2015 & False & 43 & 1 & 0 & 14.5 & F & 64 & 11.6 & \num{87173} \\
		\bottomrule
	\end{tabular}
\end{table}

For communicating these results, the five remaining weights and support points can be cast as scenarios, spanning an income range from \num{51608} to \num{100433} Euros. 
The most likely scenario, occurring with a probability of $30.6\%$, involves an income of \num{94903} Euros. 
Due to the structure of RF forecast distributions, the selected scenarios are directly associated with five actual observations occurring in our training sample.
In particular, three of the five scenarios (number 1, 4 and 5) listed in Table \ref{tab:soep_ex} involve responses of the same individual in previous years. This situation is natural since some of the regressors are nearly constant over time, so that the test point $\vx_0$ is likely to be similar to training data $\vx_i$ corresponding to the same individual. 
Using the same individual's past incomes also seems plausible from a practical perspective.\footnote{If one wanted to avoid the use of data from the same individual, one could reduce the data set to a single observation per individual, and otherwise use the same methodology.} 
Table \ref{tab:soep_ex} also yields relevant (albeit implicit) substantive information: 
For example, the fact that all five scenarios involve individuals of the same gender, family status and industry sector indicate that the RF model considers these regressors highly relevant for prediction. Conversely, the five scenarios cover all survey years present in the training data (2015-2018), indicating that time variation in incomes is less important.

\paragraph{Related Literature}
We approach the interpretability of a RF from a different angle compared to most other existing literature on the topic:
In \citeauthor{Breiman2001}'s original work on RFs, feature permutation was introduced, effectively quantifying the influence of each feature on the overall performance of the model. 
This method is still widely used and implemented in standard software packages \citep{scikit-learn}. 
\citet[Section 5]{BiauScornet2016} provide further discussion. 
This  perspective on interpretation is also commonly used for other forecasting models, such as neural networks, see e.g. \citet{Lundberg2017}. 
Visualizations of RFs have also been proposed \citep{Zhao2019, Haddouchi2019}. 
In addition, alternative notions of sparsity have been considered in the context of tree-type models.
For example, \citet{Nan2016} use the notion of sparsity in terms of feature usage in order to speed up predictions.
Several other studies consider combinations of simple prediction rules, building upon ideas in \cite{Friedman2008}.
The `node harvest' method by \cite{Meinshausen2010} is particularly interesting since it can be cast as a weighted empirical distribution, and can thus be used for probabilistic prediction. 
Node harvest first generates a large number of `nodes', each of which defines a simple subspace of the predictor space. 
It then uses a constrained optimization problem to select the nodes that are most useful for prediction. 
Although sparsity is not enforced, \citeauthor{Meinshausen2010} finds that the optimal solution often involves a small number of nodes only. 
However, this type of sparsity (with respect to nodes) is different from the type of sparsity we consider, which is with respect to the number of training sample observations being considered for prediction. 
Indeed, these two notions of sparsity seem highly complementary: 
A forecast distribution constructed via node harvest may involve a large number of training sample observations, i.e. be far from sparse in our sense. 
Conversely, a forecast distribution constructed using our proposed methodology implicitly involves many regression trees, thus being far from sparse in \citeauthor{Meinshausen2010}'s sense. 

Finally, many extensions and modifications of \citeauthor{Breiman2001}'s original RF have been proposed in the literature, typically with the aim of improving statistical performance (either empirically, or in terms of theoretical properties). 
See \cite{BiauScornet2016} for a survey, and for example \cite{BeckEtAl2023} for a forecast combination perspective, \cite{CevidEtAl2022} for a multivariate model that links RFs to kernel-based methods, and \citet{Wager2018} for an adaption to causal inference. 
By contrast, we focus on a standard, univariate RF implementation, and post-process its forecast distribution in a way that makes it easier to interpret. 
The procedure we propose can easily be applied to other RF variants (or even to prediction methods other than RFs), provided that they can be represented as discrete distributions of the type described above.

\paragraph{Roadmap of the Paper}
The remainder of this paper is organized as follows: 
Section \ref{sec:theory} introduces our methodological setup, including RFs and methods for evaluating forecast distributions. In Section \ref{sec:experiments}, we study the performance of our `Top$k$' simplification in a series of empirical experiments based on 18 data sets for which RFs have been found to perform well, as compared to deep neural network models \citep{Grinsztajn2022}. 
Our findings indicate that already for $k = \{5,10\}$, the simplified RFs can perform on a similar level compared to the full counterpart for both probabilistic and point forecasts, depending on the data set. 
Considering probabilistic forecasts, for $k = 20$, the median performance across all data sets is equivalent to the full RFs and considering $k=50$ even increases the median performance slightly. 
For point (mean) forecasts, $k=20$ even increases performance slightly.
Even though very sparse choices like $k \in \{3,5\}$ may come with a significant performance decrease, we argue that they may be worthwhile if ease of communication is a main concern. 
We further show that our results are qualitatively robust to different hyperparameter choices. 
In order to rationalize the empirical results, Section \ref{sec:theo_cons} then considers a detailed analytical example that models the weights estimated by RFs as a random draw from a Dirichlet-type distribution. 
The example also features a true vector of weights that may be either similar to, or different from, the estimated weights. 
This setup is useful to study how a simplifying approximation similar to Top$k$ affects statistical forecasting performance. 
In a nutshell, the amount of noise in the estimated weights determines whether or not the simplification comes at a high cost in terms of performance. 
Perhaps surprisingly, one can construct examples in which simplification improves performance:
this result arises when the largest weights are estimated precisely, whereas smaller weights are more noisy. 
Focusing on the largest weights then constitutes a beneficial form of shrinkage. 
When the small weights are estimated precisely, however, simplification is harmful in terms of performance.
The online supplement contains further empirical results and detailed derivations for Section \ref{sec:theo_cons}. Replication materials are available at \url{https://github.com/kosnil/simplify_rf_dist}.

\section{Methodological Setup}
\label{sec:theory}
In this section, we describe RF based forecasting as well as the forecast evaluation methods we consider.

\subsection{Forecasting Methods}
\label{subsec:theory_background}

Here we describe Random Forests and their probabilistic cousins, Quantile Regression Forests. 
We follow \citet{Lin2006} and \citet{Meinshausen2006} who emphasize the perspective of RF predictions as a weighted sum over training observations. We refer to \citet{Hastie2009} for a textbook presentation of RFs.

Our goal is to fit a univariate forecasting model. 
That is, we have some data set $\train = (\vx_i,y_i)_{i=1}^{n}$ of training set size $n$, where  $\vx_i$ is a $p$-dimensional vector of features, and $y_i \in \mathbb{R}$ is a real-valued outcome. 
We use this data set to train our model $\hat f$, such that some choice of loss function $\gL_n$ is minimized: 
\begin{equation*}
    \min_{\hat f} \gL_n(\hat f) = \frac{1}{n}\sum_i^n \gL(\hat f(\vx_i), y_i)
\end{equation*}
In a typical regression context, the function $\hat f: \mathbb{R}^p \rightarrow \mathbb{R}$ maps $\vx_i$ to the real line in order to estimate the conditional expectation functional. A popular loss function is given by squared error, with $\gL(z, y) = (y-z)^2$. Let further $\gB \subseteq \sR^p$ denote the feature space, i.e., the space in which the individual input samples $\vx_i$ exist.\\

\paragraph{Random Forests}
An RF is an ensemble of individual regression trees, each denoted by $\gT(\xi)$, where $\xi$ describes the configuration of the tree. 
At each node, a single tree greedily splits $\gB$ and rectangular subspaces thereof into two further rectangular subspaces, such that the loss $\mathcal{L}_n$ is minimized.
Each resulting subspace corresponds to a leaf $\gR_l \subseteq \gB,~l = 1,\dots,L$ where $L$ is the total number of leaves. 
Each sample $\vx_i$ can only occur in one leaf or, put differently, when dropping a sample $\vx_i$ down the tree, it can only fall into one leaf. This leaf is denoted $\ell(\vx_i, \xi) \in \{1, \ldots, L\}$ for tree $\gT(\xi)$. For a single tree $\mathcal{T}(\xi)$, a prediction $\hat \mu_\gT(\vx_0)$ for a new sample $\vx_0$ is obtained by taking the mean of all training samples within leaf $\ell(\vx_0, \xi)$. 
This can be expressed as
\begin{equation}
    \hat \mu_\gT(\vx_0) = \sum_{i=1}^{n} w_i(\vx_0,\xi) y_i,
\end{equation}
where the weight $w_i(\vx_0,\xi)$ is equal to zero for all training samples $i$ that fall into leaves other than $\ell(\vx_0, \xi)$, and is equal to one over the leaf size for all training samples that fall into $\ell(\vx_0, \xi)$: 
\begin{equation}
    w_i(\vx_0,\xi) = \frac{\ind\{\vx_i \in \gR_{\ell(\vx_0,\xi)}\}}
{\sum_{j=1}^{n}\ind\{\vx_j \in \gR_{\ell(\vx_0,\xi)}\}}. \label{eq:weight}
\end{equation}
Motivated by the lack of stability and tendency to overfit of individual trees, RFs build $B$ trees $(\mathcal{T}(\xi_b))_{b=1}^B$, based on $B$ bootstrap samples of $\train$, and consider their average prediction. 
Moreover, in each split within each tree, it is common to consider only a random subsample of $\tilde p$ out of $p$ regressors. This step aims to diversify the ensemble of trees by avoiding excessive use of the same regressors for splitting. Common choices for $\tilde p$ are $\lfloor \sqrt{p} \rfloor$ or $\lfloor\frac{p}{3}\rfloor$ \citep{Probst2019}, where $\lfloor z \rfloor$ floors the real number $z$ to the nearest integer. The RF mean prediction can thus be expressed as 
\begin{eqnarray}
    \hat \mu_{\text{RF}}(\vx_0) &=& \frac{1}{B} \sum_{b=1}^B \sum_{i=1}^{n} w_i(\vx_0,\xi_b) y_i \nonumber\\
    &=& \sum_{i=1}^{n} w_i(\vx_0) y_i, \label{eq:rf_mu_pred}
\end{eqnarray}
where
\begin{equation}
w_i(\vx_0) = \frac{1}{B}\sum_{b=1}^B  w_i(\vx_0,\xi_b) \label{eq:rf_weights}
\end{equation}
is the weight for training sample $i$, averaged across all $B$ trees. By construction, the weights $(w_i(\vx_0))_{i=1}^{n}$ are non-negative and sum to one. 
Thus, $w_i(\vx_0)$ can be interpreted as the empirically estimated probability that the new test sample observation is equal to $y_i$. \\
\paragraph{Quantile Regression Forest}
Conceptually, $\hat \mu_{\text{RF}}(\vx_0)$ is an estimate of the conditional mean $\E[Y|X=\vx_0]$. 
As described above, it is obtained as a weighted sum over all training observations. 
\citet{Meinshausen2006} extends this framework to estimating the cumulative distribution function (CDF) of $Y$, which is given by $\E[\ind(Y \leq t)|X=\vx_0] = \mathbb{P}(Y \leq t|X=\vx_0)$, where $t \in \mathbb{R}$ is a threshold value. 
The similarity to RFs becomes apparent in the last expression. Utilizing the weights from Equation \ref{eq:rf_weights}, one can approximate the CDF by the weighted mean over the binary observations $\ind(y_i \leq t)$:
\begin{equation}
    \widehat{\mathbb{P}}(Y \leq t|X=\vx_0) = \sum_{i=1}^{n} w_i(\vx_0) \ind(y_i \leq t).
    \label{eq:rf_ecdf_pred}
\end{equation}
That is, QRFs estimate the CDF of $Y$ via the weighted empirical CDF of the training sample outcomes $(y_i)_{i=1}^{n},$ using the weights $(w_i(\vx_0))_{i=1}^{n}$ produced by RFs. 
This estimator is practically appealing as it arises as a byproduct of standard RF software implementation.
Furthermore, its representation in terms of a weighted empirical CDF enables a theoretical understanding of its properties by leveraging tools from nonparametric statistics \citep{Lin2006,Meinshausen2006}.
In this paper, we consider the standard variant of QRFs which uses squared error as a criterion for finding splits (and thus growing the forest's individual trees). Various other splitting criteria have been analyzed in the literature. In particular, \cite{CevidEtAl2022} propose to use a splitting criterion based on distributional similarity. Since their RF variant retains the weighted empirical CDF representation (see their Section 2.2), our Top$k$ method can be applied to it as well. 

\subsection{Forecast Evaluation}
\label{sec:crps}

Since we generate probabilistic forecasts, we need a tool to evaluate them. 
For this, we use the Continuous Ranked Probability Score (CRPS), a strictly proper scoring rule. 
Scoring rules are loss functions for probabilistic forecasts. 
We use them in negative orientation, so that smaller scores indicate better forecasts. 
When evaluated using a proper scoring rule, a forecaster minimizes their expected score by stating what they think is the true forecast distribution. Under a strictly proper scoring rule, this minimum is unique within a suitable class of forecast distributions. 
Conceptually, strictly proper scoring rules incentivize careful and honest forecasting. 
See \cite{Gneiting2007} for a comprehensive technical treatment, and \cite{winkler1996scoring} and \cite{GneitingKatzfuss2014} for further discussion and illustration. 
The CRPS \citep{MathesonWinkler1976} is defined as
\begin{equation}
    \text{CRPS}(\hat{F}, y) = \int_{-\infty}^\infty \left(\hat{F}(z) - \ind\{z \geq y\} \right)^2 \,dz
    \label{eq:intcrps}
\end{equation}
where $\hat F$ denotes the CDF implied by the forecast distribution and $y$ denotes the true outcome. \citet[Section 4.2]{Gneiting2007} discuss connections to the concept of statistical energy \citep[e.g.][]{SzekelyGabor2017} and note that the CRPS can equivalently be expressed via its kernel representation given by $\text{CRPS}(\hat{F}, y) = \mathbb{E}_{\hat F}\big[\vert Z-y\vert \big] - \frac{1}{2}\mathbb{E}_{\hat F}\left[\vert Z-Z'\vert\right],$ where $Z$ and $Z'$ are two independent draws from $\hat F$. For various forms of $\hat F$, efficient computation methods for the CRPS are available, either via analytical expressions (if $\hat F$ is parametric) or via a representation due to \cite{LaioTamea2007} (if $\hat F$ is a discrete empirical distribution). See \cite{Jordan2019} for an overview, and \cite{Jordan2016} for details. 

The CRPS allows for very general types of forecast distributions $\hat F$. 
In particular, the forecast distribution may be discrete, that is, it need not possess a density. 
This allows for evaluating forecast distributions based on (weighted) empirical CDFs, which arise in the case of QRFs. 
In the special case that the forecast distribution is deterministic, i.e., it places point mass on a single outcome, the CRPS reduces to the Absolute Error (AE). 
Thus, numerical values of the AE and CRPS can meaningfully be compared to each other.

\citet{Jordan2019} provide an efficient implementation of the CRPS for weighted empirical distributions in their \texttt{R}-package \texttt{scoringRules}. 
Let $(y_i)_{i=1}^{n}$ denote the response values from the training data, and denote by $y_{(i)}$ their $i$th ordered value, with $y_{(1)} \le y_{(2)} \le \ldots \le y_{(n)}$. 
Furthermore, let $w_{(i)}$ denote the weight corresponding to $y_{(i)}$.
Then the CRPS for a realization $y \in \mathbb{R}$ is given by
\begin{equation}
\label{eq:crps}
\text{CRPS}(\hat F, y) = 2 \sum_{i=1}^{n} w_{(i)}(y_{(i)} - y) \left( \ind\{y < y_{(i)}\} - \left(\sum_{j=1}^iw_{(j)}\right) + \frac{w_{(i)}}{2} \right),
\end{equation}
where we dropped the dependence of $w_{(i)}$ on a vector $\vx_0$ of covariates at this point for ease of notation. 
Equation \ref{eq:crps} extends \citeauthor{Jordan2019}'s Equation 3 to the case of non-equal weights, based on their implementation in the function \texttt{crps\_sample}. In the case of sparse weights, one may omit the indices $i$ with $w_{(i)} = 0$ from the sum at (\ref{eq:crps}) in order to speed up the computation. 

Additionally to the CRPS, we also report results for the squared error (SE). 
In the present context, the SE is given by 
\begin{equation}
\text{SE}(\hat F, y) = (y - \sum_{i=1}^{n} w_i y_i)^2. \label{eq:se}
\end{equation}
Hence, the SE depends on the forecast distribution $\hat F$ via its mean $\sum_{i=1}^{n} w_i y_i$ only. 
\section{Experimental Results}
\label{sec:experiments}

In order to assess the statistical performance of the simplified forecast distributions, we conduct experiments on 18 data sets considered by \citet{Grinsztajn2022} in the context of numerical regression. The authors demonstrate that tree-based methods compare favorably to neural networks for these data sets. 
Their selection of data sets aims to represent real-world, `clean' data sets with medium size as well as heterogeneous data types and fields of applications. If deemed necessary, some basic preprocessing was applied by \citet{Grinsztajn2022}. 
Details can be found in Section 3.5 and Appendix A.1 in their paper, and in Table S3 of our online supplement. 
The data set \texttt{delays\_zurich\_transport} contains about 5.6 million data points in its original form. 
For computational reasons, we reduced the size of this data set through random subsampling, to approximately 1.1 million data points (20\% of the original observations).
We did not apply further preprocessing of any of the data sets in order to retain comparability. 
We allocate 70\% of each data set for training our models and reserve the remaining 30\% for testing. 
For each data set, we train a RF, and then evaluate its performance by computing the average CRPS and SE, as introduced in Equations \ref{eq:crps} and \ref{eq:se} over the test data set. Our main interest is in studying the impact of the parameter $k$ which governs the number of support points of the sparsified forecast distribution. We consider a grid of choices $k = 1,\ldots,50$ and denote these sparse RFs as `Top$k$'.
Our standard choice of RF hyperparameters is a combination of default values of the machine learning software packages \texttt{scikit-learn} \citep{scikit-learn} and \texttt{ranger} \citep{Wright2017} which are popular choices in the \texttt{Python} \citep{Python} and \texttt{R} \citep{R} programming languages, respectively. 
In summary, we consider a random selection of $\sqrt{p}$ out of $p$ possible features at each split point, do not impose any form of regularization in terms of tree growth restriction, and set the number of trees to 1000 in order to obtain a large and stable ensemble. These choices, which are also listed in the first row (entitled `standard') of Table S5 in the online supplement, are used for the analysis in Sections \ref{subsec:exp_prob} and \ref{subsec:exp_mean}. In Section \ref{subsec:exp_hps}, we further consider the effects of tuning hyperparameters.

\subsection{Probabilistic Forecasts}
\label{subsec:exp_prob}
Table \ref{tab:res_crps_rel} presents the CRPS for the full RF and the relative CRPS of Top$k$, for $k \in \{3, 5, 10, 20, 50\}$, compared to the full RF. For example, a relative CRPS of 1.5 indicates a 50\% larger CRPS for the Top$k$ version compared to the full RF. The three smallest values for $k$ seem especially attractive in terms of simplicity and ease of communication. 
While the two larger choices $k \in \{20, 50\}$ are less attractive in terms of simplicity, we also consider them in order to assess the trade-off between simplicity and performance. 

\vspace{.5cm}
\begin{table}[h]
    \centering
    \small
    \caption{\textbf{Results for forecast distributions.} The table reports the CRPS of the full RF as well as the CRPS for Top\{3,5,10,20,50\} relative to the full RF, i.e., $\nicefrac{\text{CRPS}_{\text{Top$k$}}}{\text{CRPS}_{\text{Full}}}$. A value smaller than 1 means that Top$k$ outperforms the full RF. 
    The last row lists the median relative CRPS across data sets.}
    \label{tab:res_crps_rel}
    \begin{tabular}{ll@{\hskip 0.2in}ccccc} 
        \toprule
                                & Absolute CRPS & \multicolumn{5}{c}{CRPS relative to Full}  \\
        Data set                & Full          & Top3  & Top5  & Top10     & Top20         & Top50 \\
        \midrule
        cpu\_act & 1.2508 & 1.21 & 1.08 & 1.00 & 0.96 & 0.95 \\
        pol & 1.4203 & 1.50 & 1.28 & 1.10 & 1.01 & 0.96 \\
        elevators & 0.0014 & 1.22 & 1.10 & 1.01 & 0.96 & 0.95 \\
        wine\_quality & 0.2565 & 1.35 & 1.20 & 1.09 & 1.02 & 0.99 \\
        Ailerons & 0.0001 & 1.23 & 1.11 & 1.01 & 0.97 & 0.96 \\
        houses & 0.1229 & 1.19 & 1.08 & 0.99 & 0.96 & 0.95 \\
        house\_16H & 0.1984 & 1.32 & 1.17 & 1.06 & 1.01 & 0.98 \\
        diamonds & 0.1320 & 1.34 & 1.21 & 1.11 & 1.05 & 1.01 \\
        Brazilian\_houses & 0.0256 & 0.88 & 0.82 & 0.78 & 0.79 & 0.85 \\
        Bike\_Sharing\_Demand & 46.0292 & 1.26 & 1.15 & 1.06 & 1.02 & 1.00 \\
        nyc-taxi-green-dec-2016 & 0.1505 & 1.39 & 1.24 & 1.12 & 1.06 & 1.02 \\
        house\_sales & 0.0995 & 1.25 & 1.14 & 1.04 & 0.99 & 0.97 \\
        sulfur & 0.0123 & 1.05 & 0.97 & 0.94 & 0.95 & 0.97 \\
        medical\_charges & 0.0376 & 1.35 & 1.21 & 1.11 & 1.05 & 1.01 \\
        MiamiHousing2016 & 0.0778 & 1.20 & 1.10 & 1.02 & 0.98 & 0.97 \\
        superconduct & 3.6341 & 1.14 & 1.08 & 1.02 & 1.00 & 0.99 \\
        yprop\_4\_1 & 0.0140 & 1.38 & 1.25 & 1.14 & 1.08 & 1.03 \\
        delays\_zurich\_transport & 1.6174 & 1.33 & 1.19 & 1.10 & 1.05 & 1.02 \\
        \midrule
        Median   & -      & 1.25   & 1.14  &  1.05  &  1.00   &   0.98  \\ 
        \bottomrule
    \end{tabular}
\end{table}

Using only $k = 3$ support points performs worse than the full RF, with a median performance cost of 25\% across data sets. 
While this result is unsurprising from a qualitative perspective, its magnitude is interesting, and gives a first indication of the performance cost of using a rather drastic simplification of the original forecast distribution. For each single data set, we find that the performance of Top$k$ improves monotonically when increasing $k$ from $3$ to $5$, from $5$ to $10$, and for all data sets but one when increasing $k$ from $10$ to $20$. 
This pattern is plausible, given that we move from a drastic simplification ($k = 3$) to less drastic versions. 
Compared to the full RF, Top5 implies a median loss increase of 14\%, whereas Top$10$ yields a median loss increase of 5\%. 
Top$20$ performs equally well as the full RF in the median. For most data sets, Top50 slightly enhances predictive performance compared to the full RF. Whether $k = 50$ support points remain worth interpreting depends on the application at hand. 
Only for a few data sets, Top50 is not sufficient to reach the performance of the full RF, but performance costs are small even for these data sets. 

In order to contextualize the magnitude of our presented results (such as Top$3$'s median CRPS increase of $25\%$ compared to the full RF), we next present results on two simple benchmark methods.\footnote{We refer to \cite{Grinsztajn2022} for detailed comparisons of RF point forecasts to other tree-based models and neural networks.}
First, we consider a deterministic point forecast which assumes that the full RF's median forecast materializes with probability one.\footnote{Formally, this forecast is characterized by the CDF $\hat F_\delta(z) = \ind(z \geq \text{med}(\hat F_{\text{Full}}(z))$, where $\text{med}(\hat F_{\text{full}})$ denotes the median implied by the CDF of the full RF's forecast distribution. 
That is, the CDF $\hat F_\delta$ is a step function with a single jump point at the median forecast of the full RF.
We choose the median functional here because the latter is the optimal point forecast under absolute error loss, to which the CRPS reduces in the case of a deterministic forecast.}

\begin{figure}[h]
    \centering
    \includegraphics[width =  1.\textwidth]{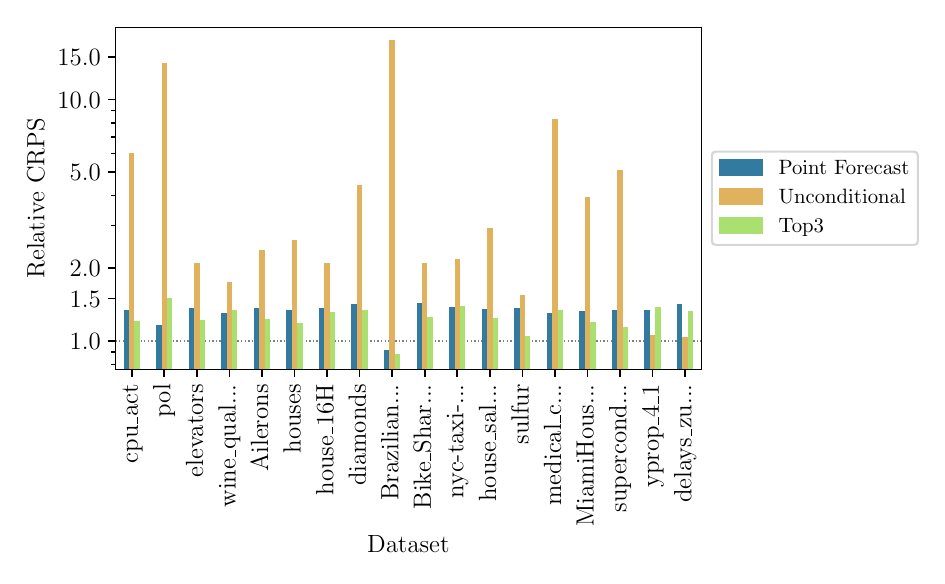} 
    \caption{\textbf{Benchmarking performance.}
    Blue bars: Relative CRPS of a deterministic point forecast (median) as a benchmark that ignores uncertainty. 
    Ochre bars: Relative CRPS of the unconditional forecast distribution, yielding a conservative benchmark that ignores features.
    Green bars: Relative CRPS of Top3, as listed in Table \ref{tab:res_crps_rel}. 
    In each case, relative CRPS results are compared to the full RF. 
    That is, a relative CRPS smaller than 1 (represented by horizontal line) indicates that the method performs better than the full RF.}
    \label{fig:maecrpsratio}
\end{figure}
This is a very optimistic point (rather than probabilistic) forecast, containing no uncertainty. 
As noted earlier, its CRPS is the same as its AE. 
We consider this benchmark in order to quantify the costs of ignoring uncertainty altogether. We clearly expect these costs to be positive, i.e., we expect the point forecast to perform worse than the full RF. Second, we use the CRPS of the unconditional empirical distribution of the response variable in the training sample. This distribution places a uniform weight of $1/n$ on each training sample observation, in contrast to the QRF weights $w(\vx_0)$ that depend on the feature vector $\vx_0$. 
The unconditional distribution is a very conservative forecast, as no information about the features is used whatsoever. 
Qualitatively, we clearly expect this forecast to perform worse than the full RF. 
Quantitatively, the difference in performance of the unconditional versus conditional forecast distributions captures the predictive content of the features \citep[see e.g.][Section 2.5]{GneitingResin2023}. 
Both benchmarks are visualized jointly with Top3 results in Figure \ref{fig:maecrpsratio}.

As expected, the relative CRPS of the point forecast (shown in blue) exceeds one for most data sets, indicating that it is generally inferior to the full RF model. 
Notably, an exception is observed for \texttt{Brazilian\_houses}, where the point forecast's relative CRPS is smaller than one. 
This result appears to be due to prediction uncertainty being very small for this data set; see Figure 13 in the online supplement for \cite{Grinsztajn2022}. 
Indeed, for this data set, the response variable seems to mostly be a linear combination of a subset of the features. Furthermore, Top3 (shown in green) performs similar to or better than the point forecast for all data sets, demonstrating the usefulness of incorporating additional uncertainty in the forecast. Note that the point forecast has access to the full RF forecast distribution, using the median of this distribution as a point forecast. By contrast, Top3 only has access to the three most important support points of the RF forecast distribution. 

The relative CRPS of the unconditional forecast distribution (displayed in ochre in Figure \ref{fig:maecrpsratio}) exceeds two for most data sets, indicating that the features are generally very useful for prediction. 
An exception occurs for the last two data sets (\texttt{yprop\_4\_1} and \texttt{delays\_zurich\_transport}).
In these cases, the RF appears unable to learn meaningful connections between the features and the target variable.
Figure 13 in the online supplement of \citet{Grinsztajn2022} supports this interpretation, reporting low predictability (in terms of low out-of-sample $R^2$) for these data sets.
In this situation, we cannot expect a Top$k$-model to perform well compared to the full RF. 
To see this, consider the stylized case of the features being entirely uninformative. 
Subsequently, the unconditional distribution (placing a weight of $1/n$ on all training sample responses) is the best possible forecast, which is in sharp contrast to Top$k$ (for which $k$ weights are non-zero and large by construction, whereas the remaining $n - k$ weights are forced to zero). 

\vspace{.5cm}
\begin{table}[h]
    \centering
    \caption{\textbf{Top$k$ weight sums.} Average sum of un-normalized Top\{3,5,10,20,50\} weights.
    The last column reports the number of observations in the training set.
    }
    \label{tab:topk_sums}
\begin{tabular}{lccccc@{\hskip 0.2in}r}
\toprule
        & \multicolumn{5}{c}{Average Sum} \\
Data set & Top3 & Top5 & Top10 & Top20 & Top50 & $n$ \\
 \midrule
cpu\_act & 0.094 & 0.135 & 0.212 & 0.318 & 0.504 & \num{5734} \\
pol & 0.073 & 0.104 & 0.159 & 0.233 & 0.359 & \num{10500} \\
elevators & 0.119 & 0.166 & 0.252 & 0.366 & 0.555 & \num{11619} \\
wine\_quality & 0.150 & 0.189 & 0.258 & 0.350 & 0.513 & \num{4547} \\
Ailerons & 0.119 & 0.168 & 0.257 & 0.374 & 0.566 & \num{9625} \\
houses & 0.143 & 0.201 & 0.304 & 0.435 & 0.634 & \num{14447} \\
house\_16H & 0.071 & 0.102 & 0.162 & 0.247 & 0.402 & \num{15948} \\
diamonds & 0.257 & 0.348 & 0.494 & 0.656 & 0.851 & \num{37758} \\
Brazilian\_houses & 0.202 & 0.278 & 0.406 & 0.558 & 0.763 & \num{7484} \\
Bike\_Sharing\_Demand & 0.248 & 0.334 & 0.473 & 0.628 & 0.821 & \num{12165} \\
nyc-taxi-green-dec-2016 & 0.172 & 0.234 & 0.337 & 0.461 & 0.642 & \num{407284} \\
house\_sales & 0.116 & 0.160 & 0.240 & 0.345 & 0.522 & \num{15129} \\
sulfur & 0.212 & 0.296 & 0.438 & 0.602 & 0.811 & \num{7056} \\
medical\_charges & 0.223 & 0.304 & 0.438 & 0.590 & 0.789 & \num{114145} \\
MiamiHousing2016 & 0.196 & 0.267 & 0.385 & 0.520 & 0.704 & \num{9752} \\
superconduct & 0.390 & 0.497 & 0.617 & 0.708 & 0.805 & \num{14884} \\
yprop\_4\_1 & 0.111 & 0.149 & 0.216 & 0.304 & 0.456 & \num{6219} \\
delays\_zurich\_transport & 0.015 & 0.024 & 0.048 & 0.095 & 0.230 & \num{765180} \\
\bottomrule
\end{tabular}
\end{table} 

In order to further study the properties of Top$k$ forecast distributions, Table \ref{tab:topk_sums} reports the average weight sums across test cases, for different values of $k$. 
The weight sums are computed before our normalization step (see Section \ref{sec:intro}) which re-scales all weight sums to one. 
For a given choice of $k$, the unnormalized weight sums can vary in magnitude, both across data sets and from test case to test case. 
By construction, the weight sums increase with $k$. 
Interestingly, many data sets yield a large weight sum for Top3, exceeding $10\%$ for all but four data sets. 
For Top50, the average weight sum exceeds $50\%$ for most data sets. 
These numbers are remarkable, given that the data sets include thousands of training samples ($n$, see rightmost column of Table \ref{tab:topk_sums}) that could potentially be used as support points for the RF forecast distributions. 
If the weights were uniform, we would hence observe Top$k$ weight sums of $k/n$. 
This is in sharp contrast to our empirical finding that a few large weights dominate for most data sets. 
\citet{Lin2006} find similar results in their work, where they consider RFs as adaptive nearest-neighbor methods and investigate the influence of the minimum number of samples per leaf. 
Figures 1c,d and 5 show few large weights for synthetic data sets.
The presence of a small number of important weights explains why the simplification pursued by Top$k$ often results in modest (if any) performance costs as compared to the full RF. 

\subsection{Mean Forecasts}
\label{subsec:exp_mean}
Let us turn our attention towards conditional mean forecasts, for which results in terms of squared error are shown in Table \ref{tab:res_mse_rel}.
We notice a similar pattern as in the probabilistic scenario: apart from Top3 outperforming the full RF for one data set (\texttt{sulfur}), Top3 performs worse than the full RF for the remaining data sets, with a maximum performance cost of 61\% for \texttt{pol}.

\vspace{.5cm}
\begin{table}[h]
    \centering
    \small
        \caption{\textbf{Results for conditional mean forecasts.} The table reports the SE of the full RF as well as the SE for Top\{3,5,10,20,50\} relative to the full RF, i.e., $\nicefrac{\text{SE}_{\text{Top$k$}}}{\text{SE}_{\text{Full}}}$. 
        The last row lists the median relative SE for each choice of $k$ across data sets.
        }
    \label{tab:res_mse_rel}
    \begin{tabular}{ll@{\hskip 0.2in}ccccc}
        \toprule
                            & Absolute SE   & \multicolumn{5}{c}{SE relative to Full}  \\
        Data set            & Full          & Top3  & Top5  & Top10     & Top20         & Top50 \\
        \midrule
        cpu\_act & \num{6.5359} & 1.11 & 0.99 & 0.93 & 0.91 & 0.91 \\
        pol & \num{38.5033} & 1.61 & 1.37 & 1.16 & 1.03 & 0.95 \\
        elevators & \num{9.22e-6} & 1.09 & 0.98 & 0.89 & 0.85 & 0.86 \\
        wine\_quality & \num{0.3485} & 1.42 & 1.24 & 1.11 & 1.03 & 1.00 \\
        Ailerons & \num{3.22e-8} & 1.15 & 1.02 & 0.94 & 0.91 & 0.91 \\
        houses & \num{0.0590} & 1.16 & 1.03 & 0.94 & 0.91 & 0.92 \\
        house\_16H & \num{0.3011} & 1.49 & 1.31 & 1.15 & 1.06 & 1.00 \\
        diamonds & \num{0.0563} & 1.37 & 1.24 & 1.12 & 1.06 & 1.02 \\
        Brazilian\_houses & \num{0.0072} & 1.05 & 1.12 & 0.99 & 0.94 & 0.94 \\
        Bike\_Sharing\_Demand & \num{10126.7983} & 1.22 & 1.11 & 1.03 & 0.99 & 0.99 \\
        nyc-taxi-green-dec-2016 & \num{0.1528} & 1.38 & 1.23 & 1.12 & 1.05 & 1.02 \\
        house\_sales & \num{0.0382} & 1.22 & 1.11 & 1.02 & 0.96 & 0.94 \\
        sulfur & 0.0015 & \num{0.69} & 0.69 & 0.73 & 0.81 & 0.90 \\
        medical\_charges & \num{0.0073} & 1.36 & 1.22 & 1.11 & 1.05 & 1.01 \\
        MiamiHousing2016 & \num{0.0243} & 1.15 & 1.05 & 0.97 & 0.94 & 0.94 \\
        superconduct & \num{87.2592} & 1.05 & 1.02 & 0.97 & 0.95 & 0.95 \\
        yprop\_4\_1 & \num{0.0010} & 1.26 & 1.16 & 1.10 & 1.05 & 1.02 \\
        delays\_zurich\_transport & \num{9.3282} & 1.33 & 1.19 & 1.10 & 1.05 & 1.02 \\
        \midrule
        Median   & -      & 1.22   & 1.12  &  1.02  &  0.98  &  0.95  \\ 
        \bottomrule
    \end{tabular}
\end{table}
This results in a median SE increase of 22\% for Top3.
Top5 still shows a 12\% median increase and for Top10, the median performance almost matches the full RF's performance. Top20 and Top50 even outperform the full RF, yielding median improvements of 2\% and 5\%, respectively. 
Compared to the results for forecast distributions (Table \ref{tab:res_crps_rel}), the results in Table \ref{tab:res_mse_rel} indicate that the performance costs of simplicity are comparatively lower in the case of mean forecasts, with median relative losses being somewhat smaller for a given value of $k$.

\subsection{Varying Hyperparameters}
\label{subsec:exp_hps}
Compared to other modeling algorithms, RFs have relatively few hyperparameters. Nevertheless, tuning its hyperparameters can improve the performance of RFs \citep{Probst2019}. 
The most important hyperparameters control the depth of each individual tree, as well as the number of randomly selected regressors considered for splitting. The depth of a tree can be restricted directly (`maximum depth') or indirectly by restricting leaf and split sizes (`minimum leaf size' and `minimum split size', respectively). 
The number of regressors considered for splitting is often denoted as `max features' or `mtry' (the latter term is used, e.g. in the \texttt{R} packages \texttt{randomForest} \citep{LiawWiener2002} and \texttt{ranger} \citep{Wright2017}). 
In what follows, we study how different hyperparameter sets influence the Top$k$ prediction and whether a hyperparameter set that optimizes the full RF is also beneficial for Top$k$.
We therefore investigate the influence of `max features' and one of the depth-regularizing hyperparameters, `minimum leaf size', on the Top$k$ approach.  
To do so, we consider a grid search for both, the full RF and Top3, with 5-fold cross-validation on the training set of each data set. 
Due to the size of \texttt{medical\_charges}, \texttt{nyc-taxi-green-dec-2016} and \texttt{delays\_zurich\_transport}, we use a validation set which contains 25\% of the training set instead. 
Further, the latter two are down-sampled to 30\% and 15\% of their original training set size, respectively.
For brevity, our analysis of hyperparameter tuning focusses on the case $k = 3$, which is the most drastic simplification we consider.
\begin{figure}[h]
    \centering
    \includegraphics[width =  1.\textwidth]{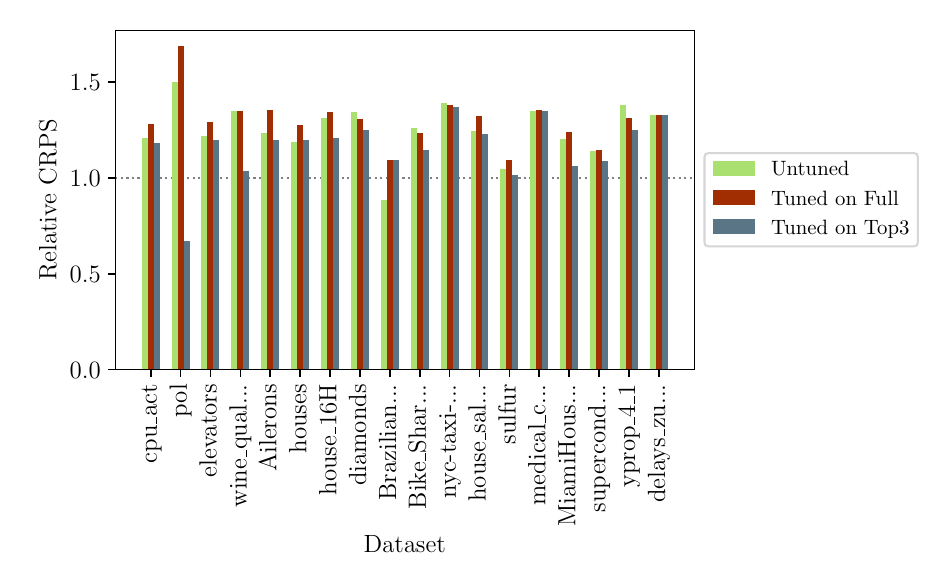} 
    \caption{\textbf{Hyperparameter tuning performance (probabilistic forecasts).}
    The figure shows the relative CRPS of Top3 compared to the full RF, for different hyperparameter settings.
    Green bars: standard hyperparameters, as listed in Table \ref{tab:res_crps_rel}.
    Red bars: hyperparameters that optimize the full RF.
    Blue bars: hyperparameters that optimize Top3. Hyperparameter tuning is based on CRPS.
    }
    \label{fig:hptuning_crps}
\end{figure}

Figure \ref{fig:hptuning_crps} visualizes the CRPS of Top3 relative to the full RF's CRPS for three different hyperparameter sets. In each case, we consider the same hyperparameter set for Top3 and the full RF. 
The green bars show the results with standard hyperparameters, as listed in Table \ref{tab:res_crps_rel}. 
The red bars indicate the relative performance with the respective hyperparameter set that optimizes the full RF, while the blue bars visualize the relative performance with the hyperparameters that optimize the CRPS of Top3.
When hyperparameters are tuned on the full RF, Top3 performance tends to slightly decrease overall. 
Across the data sets, the median relative CRPS is $1.31$, compared to $1.25$ for the standard setting. 
Conversely, if hyperparameters are tuned on Top3, the relative performance of Top3 is mostly better than in the standard version, reaching a median relative CRPS of $1.20$. 
\begin{figure}[h]
    \centering
    \includegraphics[width =  1.\textwidth]{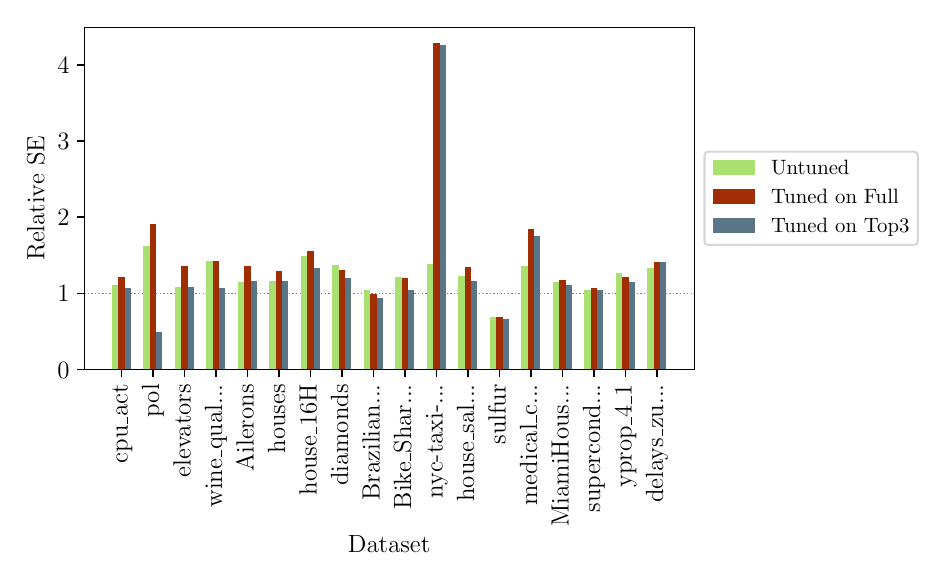} 
    \caption{\textbf{Hyperparameter tuning performance (point forecasts).}
    The figure shows the relative SE of Top3 compared to the full RF, for different hyperparameter settings.
    Green bars: standard hyperparameters, as listed in Table \ref{tab:res_crps_rel}.
    Red bars: hyperparameters that optimize the full RF.
    Blue bars: hyperparameters that optimize Top3. Hyperparameter tuning is based on SE. 
    }
    \label{fig:hptuning_se}
\end{figure}

Figure \ref{fig:hptuning_se} presents results for point forecasting performance, which are similar to the probabilistic case. 
Tuning on the full RF hurts Top3, with a median relative SE of $1.32$, compared to $1.22$ in the standard case. 
By contrast, tuning on Top3 benefits Top3, with a median relative SE of $1.13$. In a small minority of cases, tuning is not beneficial in terms of the relative loss.

Hence, despite its simplicity, Top3 can perform similar to the full RF given suitable hyperparameter choices. Overall, the relative performance of Top3 obtained under the standard setting is between the performance obtained under the two other hyperparameter settings. Furthermore, the full RF and Top3 require different sets of hyperparameters in order to perform well. Ideally, users of the Top$k$ method should thus choose hyperparameters based on the performance of Top$k$ itself, rather than the performance of the full RF. 
\section{Stylized Analytical Model}
\label{sec:theo_cons}
In this section, we construct a stylized analytical framework which helps explain our experimental findings presented in Section \ref{sec:experiments}:
For many data sets, simplified RFs perform similar to full RFs even for relatively small choices of $k$. This finding, and especially the possibility that simplified RFs may even outperform full RFs, deserves further investigation. 
Motivated by the structure of RF forecast distributions (see Section \ref{sec:theory}), we consider a model in which the true forecast distribution is discrete with support points $\rvu =(u_i)_{i=1}^n$ and corresponding (true) probabilities $\rvom^* = (\omega_i^*)_{i=1}^n$ that are positive and sum to one. 
Specifically, we let
\begin{equation}
    \omega_i^* = 
    \begin{cases} 
        \theta^*/k &~\text{if}~i \in \mathcal{I} \\ 
        (1-\theta^*)/(n-k) &~\text{if}~ i \notin \mathcal{I} \,,
    \end{cases} \label{truew}
\end{equation}
where $\mathcal{I} \subseteq \{1, 2, \ldots, n\}$ is a subset of `important' indexes with $|\mathcal{I}| = k$. 
The corresponding `important' probabilities $(\omega_i^*)_{i \in \mathcal{I}}$ sum to $\theta^* \in [0,1]$, whereas the other, `unimportant', probabilities sum to $1-\theta^*$. To justify the notion of `important' probabilities, we will focus on choices of $\theta^*$ and $k$ that satisfy $\theta^*/k > (1-\theta^*)/(n-k)$. 

In addition to the true forecast distribution just described, we consider an estimated forecast distribution that uses the same support points $\rvu$, together with possibly incorrectly estimated probabilities $\rvom = (\omega_i)_{i=1}^n$. 
We assume that the estimated probabilities can be described by the following model:
\begin{equation}
\omega_i = 
\begin{cases} 
    \theta~Z_{1,i} & ~\text{if}~i \in \mathcal{I} \\
	(1-\theta)~Z_{2,i} & ~\text{if}~i \notin \mathcal{I},
\end{cases}\label{eq:dirichlet}
\end{equation}
where $\theta \in [0,1],$ $Z_1$ is a draw from a Dirichlet distribution with $k$-dimensional parameter vector $(d_1, \ldots, d_1)$, with $d_1 > 0$, such that each element of $Z_1$ has expected value $1/k$ and variance $(k-1)/(k^2(kd_1+1))$. Similarly, $Z_2$ is a draw from another, independent Dirichlet distribution with $(n-k)$-dimensional parameter vector $(d_2, \ldots, d_2)$, where $d_2 > 0$. This means that the expected probabilities are given by
\begin{equation}
    \E[\omega_i] = 
    \begin{cases} 
        \theta/k & ~\text{if}~i \in \mathcal{I} \\
        (1-\theta)/(n-k) & ~\text{if}~i \notin \mathcal{I} \,.
    \end{cases}
\end{equation}
In the following, we assume that $2 \le k \le n-2,$ which ensures that there are at least two `important' and `unimportant' support points, respectively. This restriction ensures that weight estimation within both sets is a non-trivial problem.\footnote{If there was only one important support point, for example, the probability of this support point would necessarily be equal to $\theta$, rendering weight estimation trivial.}

Thus, if $\theta \neq \theta^*$, the forecast model's expected probabilities differ from the true ones in  Equation \ref{truew}. The parameters $d_1$ and $d_2$ represent the precision of the forecast model's probabilities around their expected values. 
Small values for $d_1, d_2$ indicate noisy probabilities, whereas large values for $d_1, d_2$ correspond to probabilities close to their expected values. This is a property of the variance of Dirichlet-distributed random variables as noted above for the case of $Z_1$. 
Conceptually, the above model provides a stylized probabilistic description of the estimated probabilities $\omega$ produced by a forecasting method like RFs. Thereby, the model does not aim to specify the mechanism by which the forecasting method generates these probabilities. 

\begin{figure}
    \centering
    \includegraphics[width=1.\textwidth]{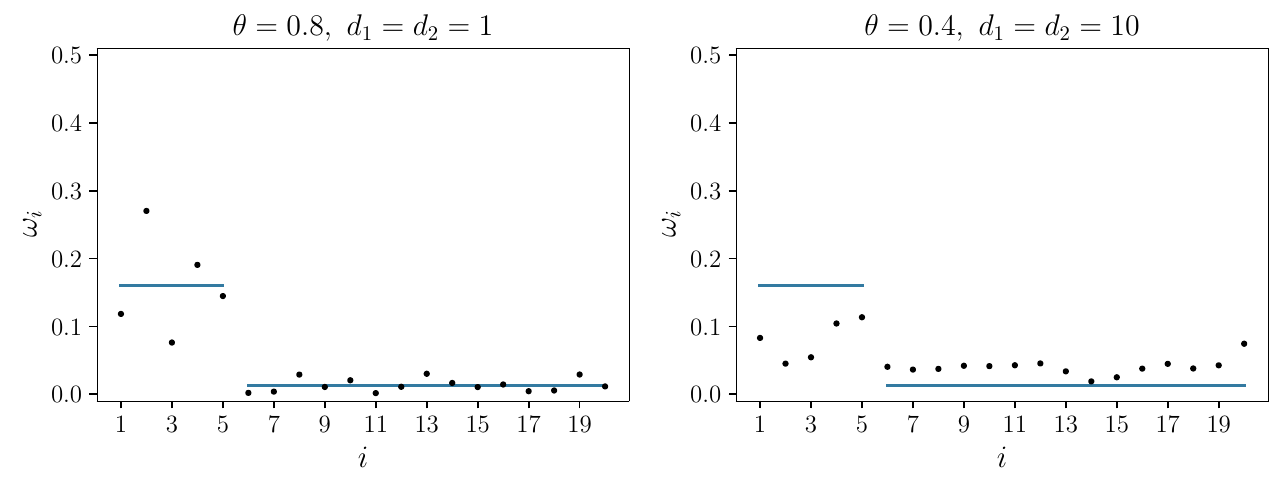}
	\caption{\textbf{Simulated probability estimates.} In both panels, we set $n = 20, k = 5$, and $\theta^* = 0.8$. The other parameters $(\theta, d_1$ and $d_2$) are as indicated in the header. Horizontal segments represent true probabilities, dots represent estimated probabilities.}
    \label{fig:simprob}
\end{figure}

Figure \ref{fig:simprob} illustrates this setup, with $n = 20$ and the important indexes given by $\mathcal{I} = \{1, 2, 3, 4, 5\}$. 
The left panel shows a situation in which the estimated probabilities are quite noisy ($d_1 = d_2 = 1)$ but are correct in expectation ($\theta = \theta^*)$. In the right panel, the estimated probabilities are less noisy ($d_1 = d_2 = 10$) but are false in expectation ($\theta = 0.4 \neq 0.8 = \theta^*$). 

In the special case that $\theta = \theta^*, d_1 \rightarrow \infty, d_2 \rightarrow \infty$, the forecast model coincides with the true model.
Furthermore, in the case $\theta = 1$, the forecast model is very similar to the `Top$k$' strategy (retaining the $k$ most important probabilities, rescaling them to sum to one, and setting all other probabilities to zero). The somewhat subtle difference between the analytical model and our practical implementation of Top$k$ is that the indexes of the important weights are fixed in the analytical model (given by the set $\mathcal{I}$), whereas they are chosen as the $k$ largest empirical weights in practice. Exact modeling of our practical procedure would seem to complicate the analysis substantially without necessarily yielding further insights. While stylized, the analytical model described above is flexible enough to cover various situations of applied interest. 
For example, the relation between `important' versus `unimportant' values of the true probabilities can be governed flexibly via the parameters $n, k$ and $\theta^*$. 
While we assume that the set $\mathcal{I}$ of important indexes is known to the forecast model, the possibility of a poor forecasting model can be represented by a value $\theta$ that differs substantially from $\theta^*$, and/or small values of $d_1, d_2$ that correspond to noisy estimates. Thus, the forecast model could even yield estimates of the `unimportant' probabilities that greatly exceed those of the `important' ones.\footnote{In practice, where the set of important indexes $\mathcal{I}$ is not known, this situation corresponds to one in which the largest empirical weights are not helpful for predicting new test sample cases.} For given support points $\rvu$ and estimated probabilities $\rvom$, the expected squared error and expected CRPS implied by the analytical framework are given by 
\begin{align}
    \label{eq:ese}
    \E[\text{SE}(\rvom, \rvu)] &= 	\sum_{i=1}^n \omega_i^*(u_i-\sum_{j=1}^n \omega_ju_j)^2,\\
    \label{eq:ecrps}
    \E[\text{CRPS}(\rvom, \rvu)] &= \sum_{i=1}^n \sum_{j=1}^n \omega_i(\omega_j^* - \omega_j/2)~|u_i-u_j|;
\end{align}
the expression for the CRPS follows from adapting the representation in Equation 2 of \cite{Jordan2019}. In both cases, the expected value is computed with respect to the discrete distribution with support points $\rvu$ and associated true probabilities $\rvom^*$.  As noted in Equation \ref{eq:dirichlet}, we cast the predicted probabilities $\rvom$ as scaled draws from two Dirichlet distributions. We further assume that the support points $\rvu$ are $n$ draws from a standard normal distribution; these are mutually independent and independent of $\rvom$. Using these assumptions, we obtain the following expressions for the (unconditionally) expected squared error and CRPS: 
\begin{align}
    \E[\text{SE}] &= \int\int \E[\text{SE}(\rvom, \rvu)]d\text{F}_\omega(\omega) dF_{\rvu}(\rvu)\nonumber\\
    &= 1 - 2\left\{ \frac{\theta^*\theta}{k} + \frac{(1-\theta^*)(1-\theta)}{n-k} \right\} \nonumber\\
    & \quad + \frac{\theta^2}{k} + \frac{(1-\theta)^2}{n-k} +  \frac{\theta^2(k-1)}{k(d_1k+1)} \nonumber \\
    & \quad + \frac{(1-\theta)^2(n-k-1)}{(n-k)(d_2(n-k)+1)}, \label{expected_se}\\[1.1em]
    \E[\text{CRPS}] &= \int\int \E[\text{CRPS}(\rvom, \rvu)]d\text{F}_\omega(\omega) dF_{\rvu}(\rvu)\nonumber\\
    &= \frac{1}{\sqrt{\pi}} \E[\text{SE}], \label{expected_crps}
\end{align}
where $F_{\rvom}$ is the distribution of the estimated probabilities that is implied by our model setup, and $F_{\rvu}$ is the joint distribution of $n$ independent standard normal variables. The proof can be found in Section C of the online supplement. The result that the expressions for $\E[\text{SE}]$ and $\E[\text{CRPS}]$ are identical up to a factor of $\sqrt{\pi}$ is a somewhat idiosyncratic implication of our model setup. 

In order to interpret the implications of these formulas, we compare a forecasting method with $\theta < 1$ (representing standard RFs) to a method with $\theta = 1$ (representing Top$k$) in the following. 

For given values of $n, k$ and $\theta^*$, both  $\E[\text{SE}]$ and $\E[\text{CRPS}]$ attain their theoretical minimum at $\theta = \theta^*, d_1 \rightarrow \infty$ and $d_2 \rightarrow \infty$.\footnote{Proof: $\frac{\partial \E[\text{SE}]}{\partial d_i} < 0$ for $i = 1, 2$; this holds for all values of $\theta, \theta^*, n, k, d_1$ and $d_2$. It is hence optimal to let $d_1, d_2$ go to infinity. 
Next consider the limiting expression of $\E[\text{SE}]$ as $d_1, d_2 \rightarrow \infty$. 
Minimizing this expression with respect to $\theta$ yields the solution $\theta = \theta^*$.} 
This result is unsurprising: Under the stated conditions, the forecast model coincides with the true model, i.e., $\omega = \omega^*$ with probability one. 
Since the squared error is strictly consistent for the mean (and, similarly, the CRPS is a strictly proper scoring rule), the true model must yield the smallest possible expected score.\footnote{While the possibility of exactly matching the true model is unrealistic in practice, the requirement that the true model perform best is conceptually plausible, and is the main idea behind forecast evaluation via proper scoring rules and related tools.} As both expected score functions are continuous in $\theta, d_1$ and $d_2$, this implies that if $\theta$ is sufficiently close to $\theta^*$, and $d_1, d_2$ are sufficiently large, then the standard approach will outperform the Top$k$ method. 

Conversely, the following conditions favor the Top$k$ method over the standard approach: 
\begin{itemize}
\setlength\itemsep{0em}
    \item $|\theta - \theta^*| > |1-\theta^*|$, i.e. the standard method's implicit assumption that $\theta^* = \theta$ is worse than Top$k$'s implicit assumption that $\theta^* = 1$
    \item $d_1$ is large, i.e. important probabilities are estimated precisely
    \item $d_2$ is small, i.e. estimates of unimportant probabilities are noisy
\end{itemize}
If these conditions, or an appropriate combination thereof, hold, then the Top$k$ approach can be expected to perform well. 
\begin{figure}
    \centering
    \includegraphics[width =  1.\textwidth]{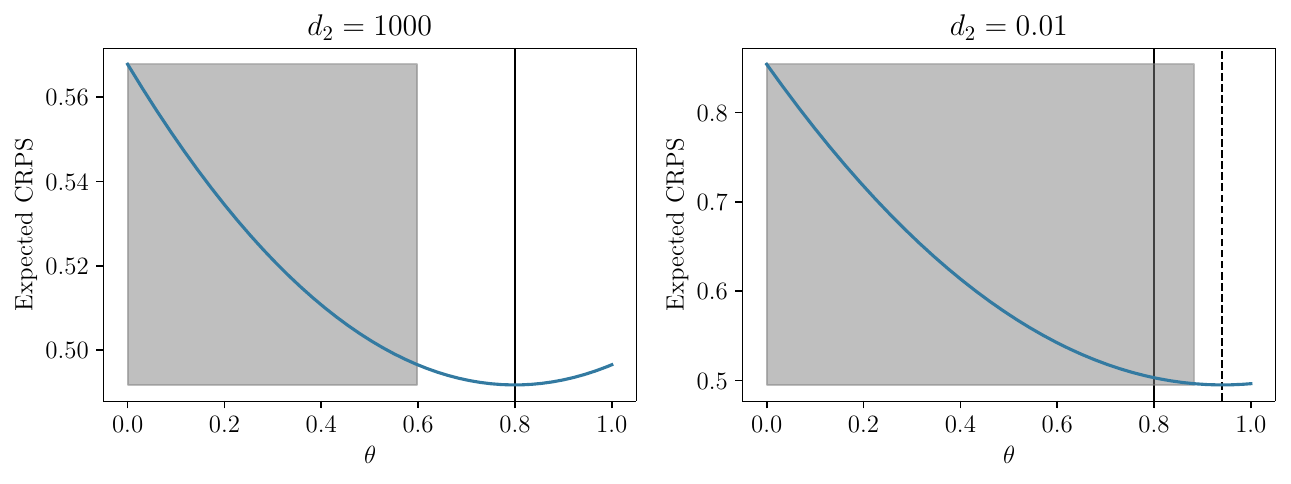} 
    \caption{\textbf{Expected CRPS as a function of $\theta$}. 
	The left panel refers to $d_2 = 1000$ (i.e., precise estimates of `unimportant' probabilities), whereas the right panel assumes $d_2 = 0.01$ (i.e., noisy estimates). The other parameters are set as follows: $n = 100, k = 5, \theta^* = 0.8, d_1 = 1000$. Solid vertical line marks $\theta^*$, dashed vertical line marks best value for $\theta$. Shaded area marks range of values for $\theta$ that perform worse than $\theta = 1$ (corresponding to the Top$k$ method).}
    \label{fig:crps_theta}
\end{figure}

Figure \ref{fig:crps_theta} illustrates the above discussion. 
In the left panel (with $d_2 = 1000$), the `unimportant' probabilities are estimated very precisely. 
Here the Top$k$ method is superior only to values $\theta \le 0.6$ that are clearly smaller than the true parameter $\theta^* = 0.8$. 
In the right panel ($d_2 = 0.01$), the estimates of the unimportant probabilities are very noisy. 
Hence it is beneficial to focus on the important probabilities which are estimated precisely (since $d_1 = 1000$). 
Accordingly, the Top$k$ method - which focuses on the important probabilities exclusively - is superior to a wide range of values for $\theta$. 
Interestingly, this range includes the true parameter $\theta = \theta^*$, i.e., the Top$k$ method can be beneficial even if the probability estimates are correct in expectation. 
\section{Discussion}
\label{sec:conclusion}

This paper has considered simplified RF forecast distributions that consist of a small number $k$ of support points, in contrast to thousands of support points (possibly equal to $n$, the size of the training set) of the original forecast distribution. 
The Top$k$ forecast distribution can be viewed as a collection of $k$ scenarios with attached probabilities. It hence simplifies communication and improves interpretability of the probabilistic forecast. 
Our empirical results in Tables \ref{tab:res_crps_rel} and \ref{tab:res_mse_rel} imply that simplified distributions using five or ten support points often attain similar performance as the original forecast distribution, while larger choices of $k$, e.g., 20 or 50, even increase performance slightly in many cases. Our analytical framework in Section \ref{sec:theo_cons} offers a theoretical rationale for these results. In particular, if small weights are estimated imprecisely by the full RF, using only the $k$ largest weights can improve performance.  Our empirical analysis further shows that when tuning hyperparameters to the target value for $k$, even $k=3$ can yield very good results.

A limitation of our proposed method is that focusing on a small number of weights is not promising if the original RF weights are close to equal. This occurs, in particular, if the regressors are entirely uninformative so that the predictive distribution for a new test point ${X} = \vx_0$ is given by the equally weighted unconditional distribution of the training responses $(y_i)_{i=1}^n$, independently of $\vx_0$. The latter distribution is consistent for the true forecast distribution as $n \rightarrow \infty$, while focusing on a finite number of $k$ weights prevents consistency. In terms of predictability, two of the 18 data sets we study are reminiscent of this situation (see Section \ref{subsec:exp_prob}). 

While we have focused on the trade-off between simplicity (as measured by $k$) and statistical forecasting performance, the optimal choice of $k$ depends on the preferences of the forecaster and the forecast users. These preferences are necessarily subjective and likely application specific. For choosing $k$ in practice, we recommend to first assess the statistical performance of various choices of $k$ for the application or data set of interest, similar to our empirical analysis in Section \ref{sec:experiments}. If ease of communication is not a concern, forecasters can simply pick the value of $k$ that optimizes (estimated) performance. Conversely, if ease of communication matters, smaller values for $k$ may be preferable even at the price of reduced statistical performance. To navigate this trade-off in practice, forecasters might interview potential users about their perceived cognitive costs of various choices of $k$. For example, \cite{AltigEtAl2022} argue that $k = 5$ resonates well with participants of an online survey on firm performance. 

Finally, the idea of simplifying predictions by sparsifying observation weights can possibly be extended to methods other than RFs. For example, the predictions of linear regression, ridge regression, and kernel representations of certain neural network models can be represented as weighted sums of training sample outcomes \citep[see e.g.][]{JacotEtAl2018,GouletKlieber2024}.

\newpage
\appendix
\FloatBarrier
\section{Details on SOEP Data}
\label{appsec:soep}

This section describes the SOEP data set used in Section 1 of the main paper in more detail.
We process the original data set with the following steps:
\begin{enumerate}
    \setlength\itemsep{0em}
    \item The variable \texttt{id}, being a unique categorical identifier for each participant, is excluded from the model.
    \item If the regressor \texttt{sector} is missing, it is imputed with the value `unknown'.
    \item The qualitative regressors \texttt{sector} and \texttt{employment} are converted into sets of dummy variables (one-hot encoding).
    \item The columns \texttt{gesund\_org} and \texttt{lebensz\_org} are excluded due to their subjective nature.
    \item The sum of variables \texttt{einkommenj1} and \texttt{einkommenj2} is selected as the target variable and referred to as `income'.
    \item Entries with missing values in any of the utilized variables are removed from the data set.
    \item Data from 2015 to 2018 is used for training, while data from 2019 is reserved for testing.
\end{enumerate}
Table \ref{apptab:soep_variables} lists all the variables utilized in the data set along their respective data type and a short description.
The both RFs, the standard and Top$k$ version, are trained using our standard hyperparameter choices, as detailed in the first row of Table \ref{apptab:hps}. Table \ref{apptab:soep_perf} reports results on forecasting performance for various choices of $k$. The SE and CRPS performance measures reported in the table are introduced in Section 2.2 of the main paper.

\vspace{.5cm}
\begin{table}[h]
    \small
    \centering
    \caption{\textbf{SOEP variables.} Description of variables in the SOEP data set, including our encoding choices after our preprocessing.}
    \label{apptab:soep_variables}
    \begin{tabular}{lll}
        \toprule
        Variable Name & Encoding & Explanation \\	
        \midrule
        id (dropped) & Categorical & Number identifying each person in the data set\\
        survey\_year & Integer & Year in which survey was taken \\
        female & Boolean &$= 1$ person is female, $= 0$ else \\
        age & Integer & Age in years \\
        n\_persons & Integer & Number of persons living in household \\
        n\_children & Integer & Number of children living in household\\
        years\_educ & Float & Years of education \\
        employed & Categorical & Employment status of person \\
        & & (6 distinct entries, description and ID given) \\
        sector & Categorical & Sector in which person is employed \\
        & & (75 distinct entries, description and ID given) \\
        income & Float & Annual income in Euros \\ 
        \bottomrule
    \end{tabular}
\end{table}

\vspace{.5cm}
\begin{table}[h]
    \centering
    \caption{\textbf{SOEP performance.} Forecasting performance on the SOEP data set of both, point and probabilistic predictions, measured with SE and CRPS, respectively. Analogous to Tables 2 and 4 in the main paper, we also report the relative losses.}
    \label{apptab:soep_perf}
    \begin{tabular}{l@{\hskip 0.3in}cccc}
        \toprule
        $k$ & SE & CRPS & Rel. SE & Rel. CRPS \\
        \midrule
        full & \num{1.55e+08} & 4822.18 & - & - \\
        \midrule
        3 & \num{1.57e+08}  & 5432.48 & 1.01 & 1.13 \\
        5 & \num{1.50e+08}  & 5090.59 & 0.97 & 1.06 \\
        10 & \num{1.47e+08} & 4874.65 & 0.95 & 1.01 \\
        20 & \num{1.48e+08} & 4801.58 & 0.96 & 1.00 \\
        50 & \num{1.51e+08} & 4786.33 & 0.97 & 0.99 \\
        \bottomrule
    \end{tabular}
\end{table}


\section{Details on Empirical Experiments}
\label{appsec:det_exp}

Here we present further details on the empirical experiments of Section 3 in the main paper. Table \ref{apptab:datasets} lists the data sets used in the experiments, following the analysis of \citet{Grinsztajn2022}. The data sets cover a wide spectrum of size, number of covariates and domains. They can be easily downloaded using the URLs listed in the table. In the case of \texttt{delays\_zurich\_transport}, we subsampled the data set to 20\% of its original size (which is shown in the table) due to computational reasons, resulting in roughly 1.1 million observations. Table \ref{apptab:possible_hps} describes the grid of hyperparameter values we consider for the analysis in Section 3.3 of the main paper, and Table \ref{apptab:hps} presents the best choices selected via cross-validation. For a given data set, we use a grid search with 5-fold cross-validation. For computational reasons, we use simplified procedures for the three largest data sets (\texttt{medical\_charges}, \texttt{nyc-taxi-green-dec-2016} and \texttt{delays\_zurich\_transport}), where we use a single holdout set which consists of 25\% of the training data. Furthermore, we subsample the \texttt{nyc-taxi-green-dec-2016} and \texttt{delays\_zurich\_transport} data sets to 30\% and 15\% of their training set size. Table \ref{apptab:hps} shows that hyperparameter choices are often the same across both loss functions (CRPS and SE), whereas differences between `Full' and `Top3' are more pronounced. Hence users of simplified RFs (such as Top3) should consider tuning hyperparameters to optimize the performance of these simplified RFs directly, instead of optimizing the performance of full RFs. 

\vspace{.5cm}
\begin{table}[h]
    \centering
    \scriptsize
    \caption{\textbf{Data sets tested.} Following \citet{Grinsztajn2022}, this table lists all tested data sets, their respective sizes (nr. of observations and regressors), name of the target variable, and URL.  
    }
    \label{apptab:datasets}
\begin{tabular}{lrrll}
\toprule
Name of Data Set                     & Number of  & Number of      & Target Variable    & URL \\
& Observations & Regressors & & \\
\midrule
cpu\_act                    & \num{8192}      & 21    & usr                   & \url{https://www.openml.org/d/44132} \\
pol                         & \num{15000}     & 26    & foo                   & \url{https://www.openml.org/d/44133} \\
elevators                   & \num{16599}     & 16    & Goal                  & \url{https://www.openml.org/d/44134} \\
wine\_quality               & \num{6497}      & 11    & quality               & \url{https://www.openml.org/d/44136} \\
Ailerons                    & \num{13750}     & 33    & goal                  & \url{https://www.openml.org/d/44137} \\
houses                      & \num{20640}     & 8     & medianhousevalue      & \url{https://www.openml.org/d/44138} \\
house\_16H                  & \num{22784}     & 16    & price                 & \url{https://www.openml.org/d/44139} \\
diamonds                    & \num{53940}     & 6     & price                 & \url{https://www.openml.org/d/44140} \\
Brazilian\_houses           & \num{10692}     & 8     & totalBRL              & \url{https://www.openml.org/d/44141} \\
Bike\_Sharing\_Demand       & \num{17379}     & 6     & count                 & \url{https://www.openml.org/d/44142} \\
nyc-taxi-green-dec-2016     & \num{581835}    & 9     & tipamount             & \url{https://www.openml.org/d/44143} \\
house\_sales                & \num{21613}     & 15    & price                 & \url{https://www.openml.org/d/44144} \\
sulfur                      & \num{10081}     & 6     & y1                    & \url{https://www.openml.org/d/44145} \\
medical\_charges            & \num{163065}    & 3     & AverageTotalPayments  & \url{https://www.openml.org/d/44146} \\
MiamiHousing2016            & \num{13932}     & 13    & SALEPRC               & \url{https://www.openml.org/d/44147} \\
superconduct                & \num{21263}     & 79    & criticaltemp          & \url{https://www.openml.org/d/44148} \\
yprop\_4\_1                 & \num{8885}      & 42    & oz252                 & \url{https://www.openml.org/d/45032} \\
delays\_zurich\_transport   & \num{5465575} $\times 0.2$   & 8     & delay    & \url{https://www.openml.org/d/45034} \\
\bottomrule
\end{tabular}
\end{table}

\vspace{.5cm}
\begin{table}[]
    \centering
    \caption{\textbf{Hyperparameter search grid.} The listed values are possible values for each hyperparameter for the hyperparameter tuning. Explanations of the hyperparameters can be found in the caption of Table \ref{apptab:hps}. This results in 44 different hyperparameter combinations. Hyperparamters indicated with an asterisk were not used for \texttt{medical\_charges} to reduce computational overhead.}
    \label{apptab:possible_hps}
    \begin{tabular}{lr}
    \toprule
    Hyperparameter & Possible Values \\
    \midrule
    min\_samples\_leaf & [1, 2, 4, 6, 8, 10, 15, 20, 30*, 40*, 50] \\ 
    max\_features & [0.333, `sqrt', 0.5, 1.0] \\
    \bottomrule
    \end{tabular}
\end{table}

\vspace{.5cm}
\begin{table}[]
    \footnotesize
    \centering
    \caption{\textbf{Hyperparameters selected via cross-validation}. The table presents the optimal hyperparameters according to 5-fold cross-validation (with exceptions for the three largest data sets, see text for details). The table lists the best choices for both hyperparameters (\texttt{max\_features} and \texttt{min\_samples\_leaf}), two loss functions (CRPS and SE) and depending on whether we consider the full RF or its simplified Top3 variant. The first row represents our standard hyperparameter choice (considered in the main paper's Sections 3.1 and 3.2) which is the same for each data set. \texttt{min\_samples\_leaf} is the minimum number of samples a leaf must contain and \texttt{max\_features} (also called \texttt{mtry} in some software packages) denotes the number of features considered in each split, where `sqrt' denotes the floored square root of the number of total features and real numbers correspond to the floored fraction of total features. Possible values for each hyperparameter are listed in Table \ref{apptab:possible_hps}. The number of trees (bootstrap iterations) is set to 1000 for all data sets, the depth remains unrestricted and the minimum number of samples required to be considered for another split is fixed to 5.} 
    \label{apptab:hps}
    \begin{tabular}{lllllrrrr}
        \toprule
            & \multicolumn{4}{c}{\texttt{max\_features}} & \multicolumn{4}{c}{\texttt{min\_samples\_leaf}} \\
        Tuned on:    & \multicolumn{2}{c}{Full} & \multicolumn{2}{c}{Top3} & \multicolumn{2}{c}{Full} & \multicolumn{2}{c}{Top3} \\
        data set & CRPS & SE & CRPS & SE & CRPS & SE & CRPS & SE \\
        \midrule
        standard & sqrt & sqrt & sqrt & sqrt & 1 & 1 & 1 & 1 \\
        \midrule
        cpu\_act & 0.5 & 0.5 & 0.333 & 0.333 & 1 & 1 & 4 & 4 \\
        pol & 0.5 & 0.5 & 0.333 & 0.333 & 1 & 1 & 20 & 20 \\
        elevators & 1.0 & 1.00 & 0.5 & 0.5 & 2 & 1 & 4 & 4 \\
        wine\_quality & 0.333 & 0.33 & 0.333 & 0.333 & 1 & 1 & 8 & 8 \\
        Ailerons & 1.0 & 1.00 & 0.5 & 0.5 & 2 & 2 & 20 & 20 \\
        houses & 1.0 & 1.00 & 0.5 & 0.5 & 2 & 2 & 4 & 4 \\
        house\_16H & 0.5 & 0.5 & sqrt & sqrt & 1 & 1 & 8 & 8 \\
        diamonds & sqrt & sqrt & 0.333 & 0.333 & 10 & 8 & 50 & 50 \\
        Brazilian\_houses & 1.0 & 1.00 & 1.0 & 0.5 & 1 & 1 & 1 & 1 \\
        Bike\_Sharing\_Demand & 0.5 & 1.00 & sqrt & sqrt & 6 & 10 & 15 & 15 \\
        nyc-taxi-green-dec-2016 & 1.0 & 1.00 & 1.0 & 1.0 & 4 & 4 & 8 & 8 \\
        house\_sales & 0.5 & 0.50 & 0.5 & 0.5 & 1 & 1 & 6 & 10 \\
        sulfur & 1.0 & sqrt & sqrt & 0.333 & 1 & 1 & 2 & 2 \\
        medical\_charges & 1.0 & 1.00 & 0.333 & 0.333 & 50 & 50 & 15 & 15 \\
        MiamiHousing2016 & 0.5 & 0.33 & sqrt & sqrt & 1 & 1 & 8 & 2 \\
        superconduct & 0.333 & 0.33 & 0.333 & sqrt & 2 & 1 & 6 & 2 \\
        yprop\_4\_1 & 0.333 & sqrt & sqrt & sqrt & 6 & 2 & 50 & 40 \\
        delays\_zurich\_transport & 0.333 & 0.33 & 0.5 & 0.5 & 50 & 50 & 4 & 8 \\
        \bottomrule
    \end{tabular}
\end{table}

\clearpage

\section{Details on the Analytical Model}
\label{appsec:exp_sc}
Here we derive the results for $\E[\text{SE}]$ and $\E[\text{CRPS}]$ stated in the main paper's Equations 14 and 15. 

\subsection{SE} 
We start with the derivation for the squared error, which is given by
\begin{equation*}
    \E[\text{SE}] = \int\int \E[\text{SE}(\rvom, \rvu)]~d\text{F}_\rvom(\rvom) dF_{\rvu}(\rvu).
\end{equation*}
First, we rewrite Equation 12 of the main paper:
\begin{align*}
    \E[\text{SE}(\rvom, \rvu)] &= \sum_{i=1}^n \rvom_i^* (u_i-\sum_{j=1}^n \rvom_j u_j)^2 \\
    &= \sum_{i=1}^n \rvom_i^*u_i^2 - 2\sum_{i=1}^n\rvom_i^*u_i\sum_{j=1}^n \rvom_ju_j \\
    & + {(\sum_{j=1}^n \rvom_ju_j)^2} \underbrace{\sum_{i=1}^n\rvom_i^*}_{=1}.
\end{align*}
We next change the order of integration and calculate the expectation with respect to the support points, i.e., we consider the expected value with respect to $\rvu$:
\begin{align*}
    \E_\rvu[\E[\text{SE}(\mathbf{\rvom}, \rvu)]] &= \int \E[\text{SE}(\mathbf{\rvom}, \rvu)]~dF_{\rvu}(\rvu)\\
    &= \E_\rvu[\sum_{i=1}^n \rvom_i^*u_i^2] - 2~ \E_\rvu[\sum_{i=1}^n \rvom_i^*u_i \sum_{j=1}^n \rvom_ju_j] + \E_\rvu[(\sum_{i=1}^n \rvom_i u_i)^2]\\
    &= 1 - 2\sum_{i=1}^n \rvom_i^* \rvom_i + \sum_{i=1}^n \rvom_i^2,
\end{align*}
where we have used the assumption that the elements of $\mathbf{u}$ are independently standard normal. Next, recall that
\begin{align*}
    \E[\rvom_i] &= 
    \begin{cases}
        \theta/k & ~\text{if}~i \in \mathcal{I} \\
        (1-\theta)/(n-k) & ~\text{if}~i \notin \mathcal{I}
    \end{cases}\\
    \Var[\rvom_i] &= 
    \begin{cases}
        \theta^2 ~ \Var[Z_{1}] & ~\text{if}~i \in \mathcal{I} \\
        (1-\theta)^2 ~ \Var[Z_{2}] & ~\text{if}~i \notin \mathcal{I}
    \end{cases}
\end{align*}
Furthermore, from the variance of a Dirichlet distributed random variable, we obtain
\begin{equation*}
    \Var[Z_{j}] = 
    \begin{cases}
        \frac{k-1}{k^2(kd_1 + 1)} & ~\text{if}~ j = 1 \\
        \frac{n-k-1}{(n-k)^2((n-k)d_2 + 1)} & ~\text{if}~ j = 2
    \end{cases}
\end{equation*}
We are now ready to calculate the expectation with respect to $\rvom$:
\begin{align*}
    \E_\rvom[\E_\rvu[\E~\text{SE}(\mathbf{\rvom}, \rvu)]] &= \int \int \E~\text{SE}(\mathbf{\rvom}, \rvu)~dF_{\rvu}(\rvu)d\text{F}_\rvom(\rvom)\\
    &= 1 - 2 \left( \sum_{i \in I} \overbrace{\rvom_i^*}^{= \frac{\theta^*}{k}} \underbrace{\E_\rvom[\rvom_i]}_{=\frac{\theta}{k}} + \sum_{i \notin I} \overbrace{\rvom_i^* }^{=\frac{1-\theta^*}{n-k}} \underbrace{\E_\rvom[\rvom_i]}_{=\frac{1-\theta}{n-k}}  \right) \\
    & \phantom{= 1}~ + (\sum_{i \in I} \E_\rvom[\rvom_i^2] + \sum_{i \notin I} \E_\rvom[\rvom_i]^2) \\
    &= 1 - 2\left\{ \frac{\theta^*\theta}{k} + \frac{(1-\theta^*)(1-\theta)}{n-k} \right\} \\
    & \phantom{= 1}~ + \frac{\theta^2}{k} + \frac{(1-\theta)^2}{n-k} +  \frac{\theta^2(k-1)}{k(kd_1+1)} + \frac{(1-\theta)^2(n-k-1)}{(n-k)((n-k)d_2+1)} \\
    & \pushQED{\qed} \qedhere \popQED
\end{align*}

\subsection{CRPS} 

We seek to evaluate the following integral:
\begin{equation*}
    \E[\text{CRPS}] = \int\int \E[\text{CRPS}(\rvom, \rvu)]~d\text{F}_\rvom(\rvom) dF_{\rvu}(\rvu),\\
\end{equation*}
where $\mathbb{E}[\text{CRPS}(\rvom, \rvu)]$ is given in the main paper's Equation 13. Note that a random variable $W \coloneqq |W_1 - W_2|$ with $W_1, W_2 \sim \mathcal{N}(0,1)$ follows a folded normal distribution with mean $\mu_Y = \nicefrac{2}{\sqrt{\pi}}$. Using this fact and Equation 13 of the main paper, the expected value with respect to $\rvu$ is given by
\begin{equation}
    \E_\rvu[\E[\text{CRPS}(\rvom, \rvu)]] = \frac{2}{\sqrt{\pi}} \sum_{i=1}^n \sum_{j \neq i} \rvom_i \left( \rvom_j^* - \frac{\rvom_j}{2} \right).\label{eq:crps_main}
\end{equation}
To simplify notation, we define $c \coloneqq \nicefrac{2}{\sqrt{\pi}}$. In order to compute the expected value of the expression in Equation 16 with respect to $\rvom$, we differentiate between four cases:

\paragraph{Case (1) $i,j \in \mathcal{I}$.}  It holds that $\Cov(\rvom_i, \rvom_j) = -\frac{\theta^2}{k^2(kd_1 + 1)}$. We hence obtain
\begin{align*}
 \sum_{i \in \mathcal{I}} \sum_{j \notin \mathcal{I}} \E_\rvom[\rvom_i] \rvom_j^* - \frac{\E_\rvom[\rvom_i\rvom_j]}{2}
    &= c(k-1) \theta \left(\frac{\theta^* }{k} - \frac{\theta d_1}{2(kd_1 + 1)} \right).
\end{align*}

\paragraph{Case (2) $i,j \notin \mathcal{I}$.} Here we have $\Cov(\rvom_i, \rvom_j) = -\frac{(1-\theta)^2}{(n-k)^2((n-k)d_2 + 1)},$ and hence
\begin{align*}
    c \sum_{i \notin \mathcal{I}} \sum_{j \notin \mathcal{I}}\left( \E_\rvom[\rvom_i] \rvom_j^* - \frac{\E_\rvom[\rvom_i\rvom_j]}{2}\right)
    &= c(n-k-1)(1-\theta) \left( \frac{(1-\theta^*)}{n-k} - \frac{(1-\theta)d_2}{2((n-k)d_2 + 1)} \right).
\end{align*}

\paragraph{Case (3) $i \in \mathcal{I},~j \notin \mathcal{I}$.} With $\Cov(\rvom_i, \rvom_j) = 0$, we obtain
\begin{align*}c \sum_{i \in \mathcal{I}} \sum_{j \notin \mathcal{I}}\left( \E_\rvom[\rvom_i] \rvom_j^* - \frac{\E_\rvom[\rvom_i\rvom_j]}{2}\right)
    &= ck(n-k) \left( \frac{\theta(1 - \theta^*)}{k(n-k)} - \frac{\theta(1-\theta)}{2k(n-k)} \right)\\
    &= c \left(\frac{\theta}{2} - \theta\theta^* + \frac{\theta^2}{2} \right).
\end{align*}

\paragraph{Case (4) $i \notin \mathcal{I},~j \in \mathcal{I}$.} It again holds that $\Cov(\rvom_i, \rvom_j) = 0$, and hence
\begin{equation*}
c \sum_{i \notin \mathcal{I}} \sum_{j \in \mathcal{I}}\left( \E_\rvom[\rvom_i] \rvom_j^* - \frac{\E_\rvom[\rvom_i\rvom_j]}{2}\right) = c \left( \theta^* -\theta\theta^* - \frac{\theta}{2} + \frac{\theta^2}{2}  \right).
\end{equation*}
Using Equation \ref{eq:crps_main}, the earlier definition $c = 2/\sqrt{\pi}$, and summarizing all four cases, we obtain 
\begin{align}
    \E[\text{CRPS}] &= \frac{2}{\sqrt{\pi}} \left\{ \theta(k-1) \left[ \frac{\theta^*}{k} - \frac{d_1\theta}{2(kd_1+1)} \right] \right\} \nonumber\\
    &+ \frac{2}{\sqrt{\pi}} \left\{ (1-\theta)(n-k-1) \left[ \frac{1-\theta^*} {n-k} - \frac{(1-\theta)d_2}{2((n-k)d_2+1)} \right] \right\} \nonumber\\
    &+ \frac{2}{\sqrt{\pi}} \left\{ \theta^* + \theta^2 - 2\theta^*\theta \right\} \\
    & \pushQED{\qed} \qedhere \popQED
    \label{crpsfinal}
\end{align}

The result that $\mathbb{E}[\text{CRPS}] = \mathbb{E}[\text{SE}]/\sqrt{\pi}$ then follows from tedious yet straightforward algebra. We are happy to provide detailed notes upon request.


\begin{thebibliography}{}

\bibitem[Abbas and Howard, 2015]{AbbasHoward2015}
Abbas, A.~E. and Howard, R.~A. (2015).
\newblock {\em Foundations of Decision Analysis}.
\newblock {Pearson}.

\bibitem[Altig et~al., 2022]{AltigEtAl2022}
Altig, D., Barrero, J.~M., Bloom, N., Davis, S.~J., Meyer, B., and Parker, N.
  (2022).
\newblock Surveying business uncertainty.
\newblock {\em {Journal of Econometrics}}, 231:282--303.

\bibitem[Beck et~al., 2023]{BeckEtAl2023}
Beck, E., Kozbur, D., and Wolf, M. (2023).
\newblock Hedging forecast combinations with an application to the random
  forest.
\newblock Preprint, arxiv:2308.15384.

\bibitem[Biau and Scornet, 2016]{BiauScornet2016}
Biau, G. and Scornet, E. (2016).
\newblock A random forest guided tour.
\newblock {\em {TEST}}, 25:197--227.

\bibitem[Breiman, 2001]{Breiman2001}
Breiman, L. (2001).
\newblock Random forests.
\newblock {\em {Machine Learning}}, 45:5--32.

\bibitem[Cevid et~al., 2022]{CevidEtAl2022}
Cevid, D., Michel, L., N{\"a}f, J., B{\"u}hlmann, P., and Meinshausen, N.
  (2022).
\newblock Distributional random forests: Heterogeneity adjustment and
  multivariate distributional regression.
\newblock {\em {Journal of Machine Learning Research}}, 23:1--79.

\bibitem[Cramer et~al., 2022]{CramerEtAl2022}
Cramer, E.~Y., Ray, E.~L., Lopez, V.~K., Bracher, J., et~al. (2022).
\newblock Evaluation of individual and ensemble probabilistic forecasts of
  {COVID}-19 mortality in the united states.
\newblock {\em Proceedings of the National Academy of Sciences},
  119:e2113561119.

\bibitem[DIW, 2022]{SoepPractice}
DIW (2022).
\newblock {SOEP-Ü}bungsdatensatz, {D}aten der {J}ahre 2015-2019.
\newblock Data set, freely available, URL:
  \url{https://doi.org/10.5684/soep.practice.v36} (last accessed: February 23,
  2025).

\bibitem[Friedman and Popescu, 2008]{Friedman2008}
Friedman, J.~H. and Popescu, B.~E. (2008).
\newblock Predictive learning via rule ensembles.
\newblock {\em {Annals of Applied Statistics}}, 2:916--954.

\bibitem[Gneiting and Katzfuss, 2014]{GneitingKatzfuss2014}
Gneiting, T. and Katzfuss, M. (2014).
\newblock Probabilistic forecasting.
\newblock {\em Annual Review of Statistics and Its Application}, 1:125--151.

\bibitem[Gneiting and Raftery, 2007]{Gneiting2007}
Gneiting, T. and Raftery, A.~E. (2007).
\newblock Strictly proper scoring rules, prediction, and estimation.
\newblock {\em Journal of the American Statistical Association}, 102:359--378.

\bibitem[Gneiting and Resin, 2023]{GneitingResin2023}
Gneiting, T. and Resin, J. (2023).
\newblock Regression diagnostics meets forecast evaluation: {C}onditional
  calibration, reliability diagrams, and coefficient of determination.
\newblock {\em {Electronic Journal of Statistics}}, 17:3226--3286.

\bibitem[Goebel et~al., 2019]{GoebelEtAl2019}
Goebel, J., Grabka, M.~M., Liebig, S., Kroh, M., Richter, D., Schr{\"o}der, C.,
  and Schupp, J. (2019).
\newblock The {G}erman socio-economic panel ({SOEP}).
\newblock {\em {Jahrb{\"u}cher f{\"u}r National{\"o}konomie und Statistik}},
  239:345--360.

\bibitem[Goulet~Coulombe et~al., 2024]{GouletKlieber2024}
Goulet~Coulombe, P., G{\"o}bel, M., and Klieber, K. (2024).
\newblock Dual interpretation of machine learning forecasts.
\newblock Preprint, available at \url{https://arxiv.org/abs/2412.13076}.

\bibitem[Grinsztajn et~al., 2022]{Grinsztajn2022}
Grinsztajn, L., Oyallon, E., and Varoquaux, G. (2022).
\newblock Why do tree-based models still outperform deep learning on typical
  tabular data?
\newblock In {\em {Proceedings of the 36th International Conference on Neural
  Information Processing Systems}}, pages 507--520. {Curran Associates Inc.}

\bibitem[Haddouchi and Berrado, 2019]{Haddouchi2019}
Haddouchi, M. and Berrado, A. (2019).
\newblock A survey of methods and tools used for interpreting random forest.
\newblock In {\em {1st International Conference on Smart Systems and Data
  Science (ICSSD)}}, pages 1--6.

\bibitem[Hastie et~al., 2009]{Hastie2009}
Hastie, T., Tibshirani, R., and Friedman, J. (2009).
\newblock {\em The Elements of Statistical Learning: Data Mining, Inference,
  and Prediction}.
\newblock {Springer}, 2 edition.

\bibitem[Jacot et~al., 2018]{JacotEtAl2018}
Jacot, A., Gabriel, F., and Hongler, C. (2018).
\newblock Neural tangent kernel: Convergence and generalization in neural
  networks.
\newblock In Bengio, S., Wallach, H., Larochelle, H., Grauman, K.,
  Cesa-Bianchi, N., and Garnett, R., editors, {\em Advances in Neural
  Information Processing Systems}, volume~31. Curran Associates, Inc.

\bibitem[Jordan et~al., 2019]{Jordan2019}
Jordan, A., Krüger, F., and Lerch, S. (2019).
\newblock Evaluating probabilistic forecasts with scoring{R}ules.
\newblock {\em {Journal of Statistical Software}}, 90:1--37.

\bibitem[Jordan, 2016]{Jordan2016}
Jordan, A.~I. (2016).
\newblock {\em Facets of Forecast Evaluation}.
\newblock PhD thesis, Karlsruher Institut f{\"u}r Technologie, 2016.

\bibitem[Kr{\"u}ger et~al., 2017]{KruegerEtAl2017}
Kr{\"u}ger, F., Clark, T.~E., and Ravazzolo, F. (2017).
\newblock Using entropic tilting to combine {BVAR} forecasts with external
  nowcasts.
\newblock {\em {Journal of Business \& Economic Statistics}}, 35:470--485.

\bibitem[Laio and Tamea, 2007]{LaioTamea2007}
Laio, F. and Tamea, S. (2007).
\newblock Verification tools for probabilistic forecasts of continuous
  hydrological variables.
\newblock {\em Hydrology and Earth System Sciences}, 11:1267--1277.

\bibitem[Liaw and Wiener, 2002]{LiawWiener2002}
Liaw, A. and Wiener, M. (2002).
\newblock Classification and regression by random{F}orest.
\newblock {\em {R News}}, 2:18--22.

\bibitem[Lin and Jeon, 2006]{Lin2006}
Lin, Y. and Jeon, Y. (2006).
\newblock Random forests and adaptive nearest neighbors.
\newblock {\em {Journal of the American Statistical Association}},
  101:578--590.

\bibitem[Lundberg and Lee, 2017]{Lundberg2017}
Lundberg, S.~M. and Lee, S.-I. (2017).
\newblock A unified approach to interpreting model predictions.
\newblock In {\em {Proceedings of the 31st International Conference on Neural
  Information Processing Systems}}, pages 4768--4777. Curran Associates Inc.

\bibitem[Matheson and Winkler, 1976]{MathesonWinkler1976}
Matheson, J.~E. and Winkler, R.~L. (1976).
\newblock Scoring rules for continuous probability distributions.
\newblock {\em {Management Science}}, 22:1087--1096.

\bibitem[Meinshausen, 2006]{Meinshausen2006}
Meinshausen, N. (2006).
\newblock Quantile regression forests.
\newblock {\em {Journal of Machine Learning Research}}, 7:983--999.

\bibitem[Meinshausen, 2010]{Meinshausen2010}
Meinshausen, N. (2010).
\newblock Node harvest.
\newblock {\em {Annals of Applied Statistics}}, 4:2049--2072.

\bibitem[Nan et~al., 2016]{Nan2016}
Nan, F., Wang, J., and Saligrama, V. (2016).
\newblock Pruning random forests for prediction on a budget.
\newblock In {\em {Proceedings of the 30th International Conference on Neural
  Information Processing Systems}}, pages 2342--2350. Curran Associates Inc.

\bibitem[Pedregosa et~al., 2011]{scikit-learn}
Pedregosa, F., Varoquaux, G., Gramfort, A., Michel, V., Thirion, B., Grisel,
  O., Blondel, M., et~al. (2011).
\newblock Scikit-learn: Machine learning in {P}ython.
\newblock {\em {Journal of Machine Learning Research}}, 12:2825--2830.

\bibitem[Probst et~al., 2019]{Probst2019}
Probst, P., Wright, M.~N., and Boulesteix, A.-L. (2019).
\newblock Hyperparameters and tuning strategies for random forest.
\newblock {\em {WIREs Data Mining and Knowledge Discovery}}, 9:e1301.

\bibitem[{R Core Team}, 2022]{R}
{R Core Team} (2022).
\newblock {\em R: A Language and Environment for Statistical Computing}.
\newblock R Foundation for Statistical Computing, Vienna, Austria.
\newblock URL: \url{https://www.R-project.org/} (last accessed: June 13, 2025).

\bibitem[Raftery, 2016]{Raftery2016}
Raftery, A.~E. (2016).
\newblock Use and communication of probabilistic forecasts.
\newblock {\em {Statistical Analysis and Data Mining: The ASA Data Science
  Journal}}, 9:397--410.

\bibitem[Rasp and Lerch, 2018]{RaspLerch2018}
Rasp, S. and Lerch, S. (2018).
\newblock Neural networks for postprocessing ensemble weather forecasts.
\newblock {\em {Monthly Weather Review}}, 146:3885--3900.

\bibitem[Sz{\'e}kely and Rizzo, 2017]{SzekelyGabor2017}
Sz{\'e}kely, G.~J. and Rizzo, M.~L. (2017).
\newblock The energy of data.
\newblock {\em Annual Review of Statistics and Its Application}, 4:447--479.

\bibitem[Taieb et~al., 2021]{TaiebEtAl2021}
Taieb, S.~B., Taylor, J.~W., and Hyndman, R.~J. (2021).
\newblock Hierarchical probabilistic forecasting of electricity demand with
  smart meter data.
\newblock {\em {Journal of the American Statistical Association}}, 116:27--43.

\bibitem[van Rossum et~al., 2011]{Python}
van Rossum, G. et~al. (2011).
\newblock {\em Python programming language}.
\newblock URL: \url{https://www.python.org} (last accessed: June 13, 2025).

\bibitem[Wager and Athey, 2018]{Wager2018}
Wager, S. and Athey, S. (2018).
\newblock Estimation and inference of heterogeneous treatment effects using
  random forests.
\newblock {\em {Journal of the American Statistical Association}},
  113:1228--1242.

\bibitem[Winkler, 1996]{winkler1996scoring}
Winkler, R.~L. (1996).
\newblock Scoring rules and the evaluation of probabilities.
\newblock {\em {TEST}}, 5:1--26.

\bibitem[Wright and Ziegler, 2017]{Wright2017}
Wright, M.~N. and Ziegler, A. (2017).
\newblock ranger: A fast implementation of random forests for high dimensional
  data in {c++} and {R}.
\newblock {\em {Journal of Statistical Software}}, 77:1--17.

\bibitem[Zhao et~al., 2019]{Zhao2019}
Zhao, X., Wu, Y., Lee, D.~L., and Cui, W. (2019).
\newblock {IF}orest: Interpreting random forests via visual analytics.
\newblock {\em {IEEE Transactions on Visualization and Computer Graphics}},
  25:407--416.

\end{thebibliography}
\end{document}